\documentclass[preprint]{aastex63}
\usepackage[utf8]{inputenc}

\usepackage{xcolor}
\usepackage{hyperref}
\usepackage[caption=false]{subfig}

\usepackage[mathscr]{euscript}

\received{}
\revised{}
\accepted{}
\submitjournal{ApJS}

\shorttitle{PHOEBE V: The Inverse Problem}
\shortauthors{Conroy et al.}

\begin{document}

\title{Physics of Eclipsing Binaries. V. General Framework for Solving the Inverse Problem}

\correspondingauthor{Kyle Conroy}
\email{kyle.conroy@villanova.edu}

\author[0000-0002-5442-8550]{Kyle E.~Conroy}
\affiliation{Department of Astrophysics and Planetary Science,
Villanova University,
800 East Lancaster Avenue,
Villanova, 
PA 19085, 
USA}

\author[0000-0002-9739-8371]{Angela Kochoska}
\affiliation{Department of Astrophysics and Planetary Science,
Villanova University,
800 East Lancaster Avenue,
Villanova, 
PA 19085, 
USA}

\author[0000-0003-3244-5357]{Daniel Hey}
\affiliation{School of Physics, Sydney Institute for Astronomy (SIfA) \\
The University of Sydney, NSW 2006, Australia}
\affiliation{Stellar Astrophysics Centre, Department of Physics and Astronomy \\
Aarhus University, DK-8000 Aarhus C, Denmark}

\author[0000-0002-1355-5860]{Herbert Pablo}
\affiliation{American Association of Variable Star Observers, 49 Bay State Road, Cambridge, MA 02138, USA}

\author[0000-0001-5473-856X]{Kelly M.~Hambleton}
\affiliation{Department of Astrophysics and Planetary Science,
Villanova University,
800 East Lancaster Avenue,
Villanova, 
PA 19085, 
USA}

\author[0000-0003-3947-5946]{David Jones}
\affiliation{Insituto de Astrof\'isica de Canarias,
E-38205 La Laguna,
Tenerife, 
Spain}
\affiliation{Departamento de Astrof\'isica, 
Universidad de La Laguna,
E-38206 La Laguna,
Tenerife, 
Spain}

\author[0000-0003-4560-7925]{Joseph Giammarco}
\affiliation{Eastern University, Dept.~of Astronomy and Physics, 1300 Eagle Rd, St.~Davids, PA 19087}

\author[0000-0001-6566-7568]{Michael Abdul-Masih}
\affiliation{Institute of Astronomy, KU Leuven, Celestijnenlaan 200 D, 3001, Leuven, Belgium}

\author[0000-0002-1913-0281]{Andrej Pr\v{s}a}
\affiliation{Department of Astrophysics and Planetary Science,
Villanova University,
800 East Lancaster Avenue,
Villanova, 
PA 19085, 
USA}

\begin{abstract}

PHOEBE 2 is a Python package for modeling the observables of eclipsing star systems, but until now has focused entirely on the forward-model -- that is, generating a synthetic model given fixed values of a large number of parameters describing the system and the observations.  The inverse problem, obtaining orbital and stellar parameters given observational data, is more complicated and computationally expensive as it requires generating a large set of forward-models to determine which set of parameters and uncertainties best represent the available observational data. The process of determining the best solution and also of obtaining reliable and robust uncertainties on those parameters often requires the use of multiple algorithms, including both optimizers and samplers.  Furthermore, the forward-model of PHOEBE has been designed to be as physically robust as possible, but is computationally expensive compared to other codes.  It is useful, therefore, to use whichever code is most efficient given the reasonable assumptions for a specific system, but learning the intricacies of multiple codes presents a barrier to doing this in practice.  Here we present the 2.3 release of PHOEBE (publicly available from \url{http://phoebe-project.org}) which introduces a general framework for defining and handling distributions on parameters, and utilizing multiple different estimation, optimization, and sampling algorithms. The presented framework supports multiple forward-models, including the robust model built into PHOEBE itself.

\end{abstract}

\section{Introduction}

Eclipsing binary systems provide the ability to determine absolute fundamental parameters of stellar components, calibrate stellar evolution models, measure the orientation and dynamical changes in orbits, and derive distances without the need for calibration.  For benchmark-grade eclipsing binaries, a precision in fundamental stellar parameters (masses and radii) of a few percent can be obtained \citep{torres2010}.  All these scientific goals can only be achieved through solving the inverse problem: models are created to simulate synthetic observables given a set of input parameters, and the input parameters can then be varied to determine which values result in synthetic observables that best represent the observations.  The parameter space, however, is complex and often not smooth, making this task expensive and time-consuming.

PHOEBE is one of many modeling codes which operates by discretizing the gravitationally distorted surface of each stellar component, populating the discretized grid with local quantities from atmosphere tables, and integrating over visible elements at any given time in the orbit to synthesize model observables.  Until now, all releases since the 2.0 release  \citep{prsa2016, horvat2018, jones2020} have been focused on adding advanced physics and precision to the forward-model and leaving the inverse problem to the user.

Due to the complexity of the parameter space, different combinations of observations available, and the wide-range of system morphologies and relevant physics effects, each individual system often requires a manual treatment in order to obtain the most accurate and precise model and uncertainties possible.  Furthermore, it is often necessary to utilize multiple algorithms in order to increase the chance of determining the optimal solution and produce reliable and realistic uncertainties on the modeled parameters.  Having to ``wrap'' several algorithms, each with its own interface, around the forward-model can present a steep learning curve and effort-overhead to solving these systems.  To address this, PHOEBE now includes a common interface to interact with a variety of algorithms, some off-the-shelf and others custom built, that are useful for solving the inverse problem.

To complicate matters, the available forward-model codes each have their own interfaces and often different parameterization, making it even more difficult to compare the forward-models or choose the code most appropriate for a given scenario.  PHOEBE 2 has been designed and developed with precision, robustness, and flexibility as the main priorities -- but this comes at the cost of computational efficiency.  In many cases, other codes may be just as capable of modeling a given system but at a fraction of the time cost.  Learning the intricacies of each of these codes presents yet another learning curve and overhead.  PHOEBE now incorporates several other publicly available codes as the forward-model, converting parameterization and units for both inputs and outputs as necessary.  This brings several large benefits, including the abilities to: easily compare the synthetic models created by different codes, choose the code most appropriate and efficient for a specific system, and transition to more expensive but precise models throughout solving the inverse problem.

Here we outline a general framework for solving the inverse problem with PHOEBE.  We divide the available solver algorithms incorporated in this PHOEBE release into three categories: ``estimators'', ``optimizers'', and ``samplers''.  Section \ref{sec:estimators} details the  ``estimators'' now available in PHOEBE that estimate system parameters from observations alone, without the need for a forward-model.  These estimators are designed to cheaply determine the parameter space of a particular system given any of the data available.  Section \ref{sec:merit_function} then discusses the merit function from the available forward-models.  This merit function is then used for both ``optimizers'' (Section \ref{sec:optimizers}) and ``samplers'' (Section \ref{sec:samplers}) to find the optimal solution and determine robust posteriors (Section \ref{sec:posteriors}).  Section \ref{sec:ui} then presents the new user interface introduced in this release, allowing easy access to most of the inverse-problem functionality from either a web browser or a dedicated desktop client.  See Appendix \ref{app:symbols} for descriptions of the symbols used for system parameters throughout the paper.

This paper is accompanied by the 2.3 release of PHOEBE, available from \url{http://phoebe-project.org} and \url{https://github.com/phoebe-project/phoebe2}.  Scripts to reproduce all figures in this paper are available at \url{http://phoebe-project.org/publications/2020Conroy+}.

\section{Estimators}\label{sec:estimators}

It is often useful and computationally prudent to estimate the values of as many parameters as possible without having to compute multiple physical models. The 2.3 release of PHOEBE incorporates several algorithms which act directly on the observations themselves, with the goal to provide fast initial solutions for a subset of the relevant system parameters without significant user supervision, including:

\begin{description}
    \item[RV periodogram, LC periodogram] computes the periodogram for light curves or radial velocity curves, respectively, and proposes an orbital period, $P_\mathrm{orb}$;
    \item[RV geometry] fits a simple Keplerian orbit to propose values for: $t_{0, \mathrm{supconj}}$, $e$, $\omega_0$, $v_\gamma$, and either $q$ and the projected orbital semi-major axis, $a_\mathrm{orb} \sin i$, or the projected per-component semi-major axes, $a_\mathrm{comp} \sin i$ ;
    \item[LC geometry] estimates phases of eclipse mimima, ingress, and egress and uses that information to propose values for: $t_{0, \mathrm{supconj}}$, $e$, and $\omega_0$; and
    \item[EBAI] uses a trained artificial neural network on the input light curves to propose values for: $t_{0, \mathrm{supconj}}$, $T_\mathrm{eff, 2} / T_\mathrm{eff, 1}$, $(R_\mathrm{equiv, 1} + R_\mathrm{equiv, 2}) / a_\textrm{orb}$, $e \sin \omega_0$, $e \cos \omega_0$, and $i$.
\end{description}

For all estimators that work in phase-space (as opposed to time-space), the input data can optionally be binned to maintain the efficiency of the estimators on large data sets.

\subsection{Periodograms}

The most common observational data sets used when modeling eclipsing star systems are light curves and radial velocity curves.  For the majority of eclipsing systems (those that do not exhibit significant time-dependent changes), these data can be phase-folded on the orbital period of the system.  If the period is well known and the orbital parameters do not change between successive orbits, then the system can be sampled in phase-space instead.

PHOEBE 2.3 contains wrappers around two common periodograms -- Box Least Squares (BLS, for light curves) and Lomb Scargle (for light curves or radial velocities) -- as implemented by \texttt{astropy.timeseries}\footnote{requires astropy v3.2 or later.}\citep{2013A&A...558A..33A}.  The wrapper takes any number of light curves or radial velocities as input, in addition to several of the advanced options from astropy, including: automatic or manual sampling in period/frequency, the option to change the objective function for BLS, and the ability to set the proposed eclipse durations for BLS.  If running on multiple light curves, each light curve is normalized by dividing by either the median or maximum flux value before sending to the periodogram algorithm.  Radial velocities across all requested data sets are combined and then normalized by the absolute maximum value for the primary and secondary star independently, with the secondary then mirrored. After running the wrapper, the periodogram itself is returned and can be plotted, and the peak-period is proposed for adoption.  Especially for near-circular systems with similar eclipses, these algorithms often find a strong signal at aliases of the true orbital period, so an easy interface is provided for adopting any factor of the proposed peak period.

\subsection{Radial Velocity Geometry}\label{sec:rv_geometry}

Analytical radial velocities (RVs) are used as an estimator for the eccentricity ($e$), argument of periastron ($\omega$), systemic velocity ($v_{\gamma}$), and the time of superior conjunction ($t_{0, \mathrm{supconj}}$), as well as the mass ratio ($q$) and projected orbital semi-major axis ($a_\mathrm{orb} \sin i$) in the case where RVs are provided for both components or the single-component projected semi-major axis ($a_\mathrm{comp, i} \sin i$) in the case where RVs are only provided for either one of the stellar components. All estimates are based on the per-component radial velocity equations as a function of the true anomaly ($\vartheta$) \citep{prsa2018}:

\begin{eqnarray}\label{rvs_eq}
RV_1 (\vartheta) & = & \frac{2\pi a_1 \sin i}{P_\mathrm{orb}\sqrt{1-e^2}}[e\cos\omega + \cos(\omega + \vartheta)] + v_{\gamma} \nonumber \\
 & = & \frac{2\pi q a \sin i}{P_\mathrm{orb}(1+q)\sqrt{1-e^2}}[e\cos\omega + \cos(\omega + \vartheta)] + v_{\gamma}, \\
RV_2 (\vartheta) & = & - \frac{2\pi a_2 \sin i}{P_\mathrm{orb}\sqrt{1-e^2}}[e\cos\omega + \cos(\omega + \vartheta)] + v_{\gamma} \nonumber \\
& = & - \frac{2\pi a \sin i}{P_\mathrm{orb}(1+q)\sqrt{1-e^2}}[e\cos\omega + \cos(\omega + \vartheta)] + v_{\gamma}.
\end{eqnarray}

We assume that the period ($P_\mathrm{orb}$) is known and use the phase-folded radial velocity curves as input. The data is smoothed with a low-pass Savitzky-Golay filter to remove high frequency noise.

The first two parameters that are estimated in parallel are the mass ratio and systemic velocity. From Eq.~\ref{rvs_eq} we can derive analytical expressions that hold at any point in the orbit for both parameters:
        \begin{equation}
            v_{\gamma} = \frac{RV_1 (\vartheta) + q RV_2 (\vartheta)}{1+q},
        \end{equation}
        \begin{equation}
            q = \frac{RV_1 (\vartheta) - v_{\gamma}}{-RV_2 (\vartheta) + v_{\gamma}}.
        \end{equation}
As we can see, $v_{\gamma}$ and $q$ appear in both expressions, which means we cannot uniquely determine them independently of each other. For that purpose we compute them iteratively, starting from a crude estimate of $q$ as the ratio of primary and secondary RV. To maximize the efficiency, we only estimate $q$ and $v_{\gamma}$ in the RV points that correspond to the maximum primary RV.  If only one RV is available, $q$ cannot be reliably estimated and in this case only $v_{\gamma}$ is estimated as the midpoint of the available RV.

With $q$ and $v_{\gamma}$ estimated, we can compute the projected semi-major axis, $a\sin i$. We first estimate the semi-amplitudes from the available observations in each RV:
    \begin{equation}
        K_i = 0.5 [\mathrm{max}(RV_i-v_{\gamma}) - \mathrm{min}(RV_i - v_{\gamma})]
    \end{equation}
    where $i = 1, 2$ refers to the primary and secondary component, respectively. The corresponding projected semi-major axes are then:
    \begin{equation}
        a_i\sin i = \frac{K_i P_\mathrm{orb}}{2\pi} \sqrt{1-e^2}
    \end{equation}
    Initially, we assume $e=0$. If both RVs are available, $a\sin i = a_1\sin i + a_2\sin i$.

Once $q$, $v_{\gamma}$ and $a\sin i$ are estimated, we fix them in the analytical radial velocities and fit for eccentricity and argument of periastron with a least-squares algorithm. We compute several solutions by iterating over a small grid of starting points with the combinations $e_0 = [0, 0.4]$ and $\omega_0 = [0, \pi/2, \pi]$ which ensures the entire feasible parameter space is covered and we iterate over the least squares solution until the values converge within a tolerance or the maximum number of iterations is reached. After each iteration, the value of $a\sin i$ is recomputed with the updated eccentricity $e$ and the solution is used as an initial point in the next iteration.
    
Finally, the phase of superior conjunction is estimated as the phase at which $v = v_{\gamma}$ and then converted to the time of superior conjunction, $t_{0, \mathrm{supconj}}$.

\begin{figure}
    \centering
    \includegraphics[width=0.5\textwidth]{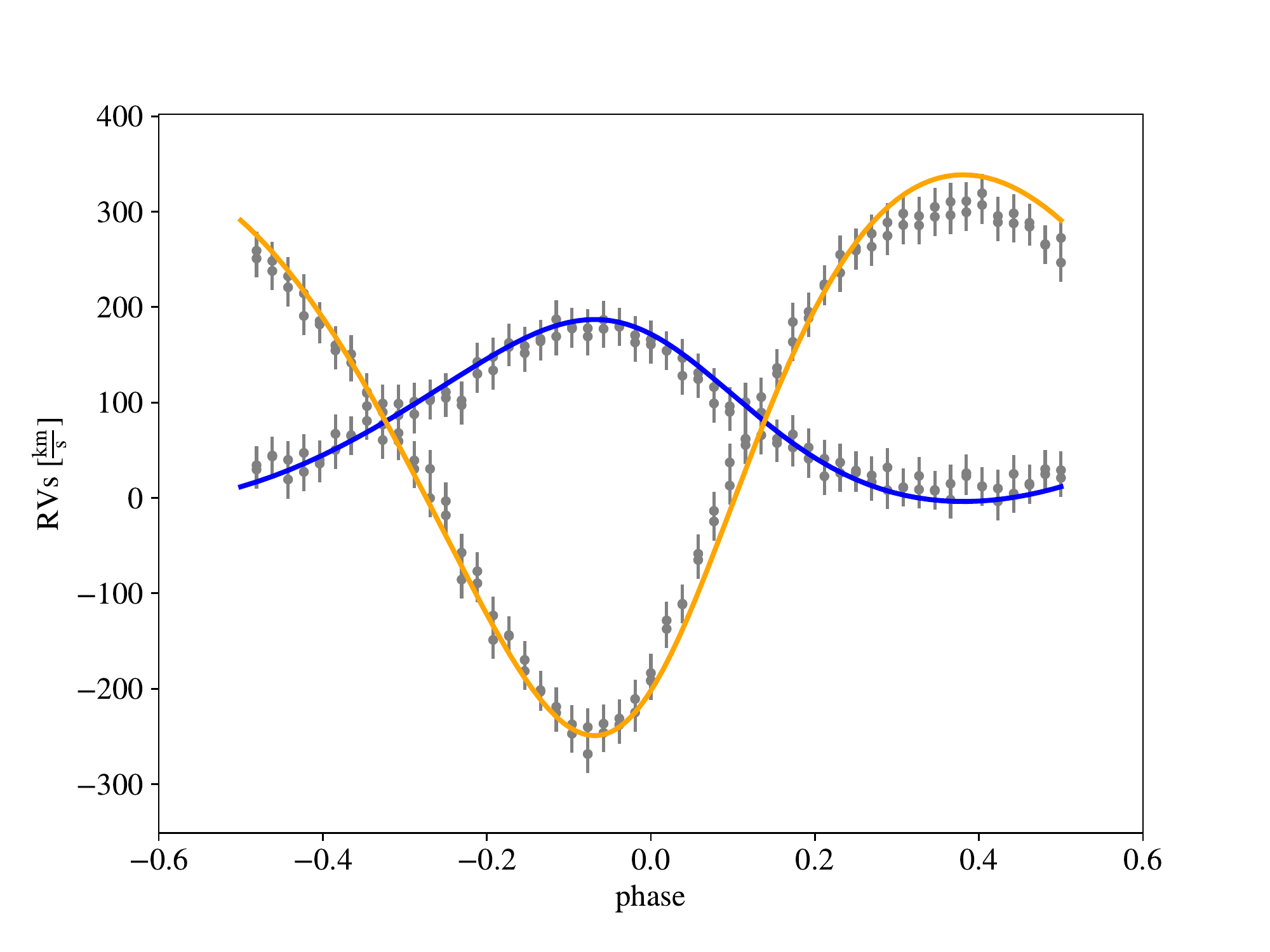} \\
    \caption{Analytic Keplerian radial velocity model fitted to a synthetic set of observations with estimated values for eccentricity ($e$), argument of periastron ($\omega_0$), mass ratio ($q$), projected semi-major axis ($a_\mathrm{orb} \sin i$), and the time of superior conjunction ($t_{0,\mathrm{supconj}}$) from only the phased radial velocity observations.}
    \label{fig:rv_geometry}
\end{figure}

An example of the performance of the estimator is given in Figure \ref{fig:rv_geometry}.

\subsection{Light Curve Geometry}\label{sec:lc_geometry}

PHOEBE 2.3 introduces an estimator based on fitting a two-Gaussian model to a phased light curve to evaluate the eccentricity and argument of periastron, as the two best-constrained parameters from light curve geometry. For a detailed description of the model refer to \citet{mowlavi2017}. Here, we will focus on its implementation in PHOEBE.

\begin{figure}
    \centering
    \includegraphics[width=0.5\textwidth]{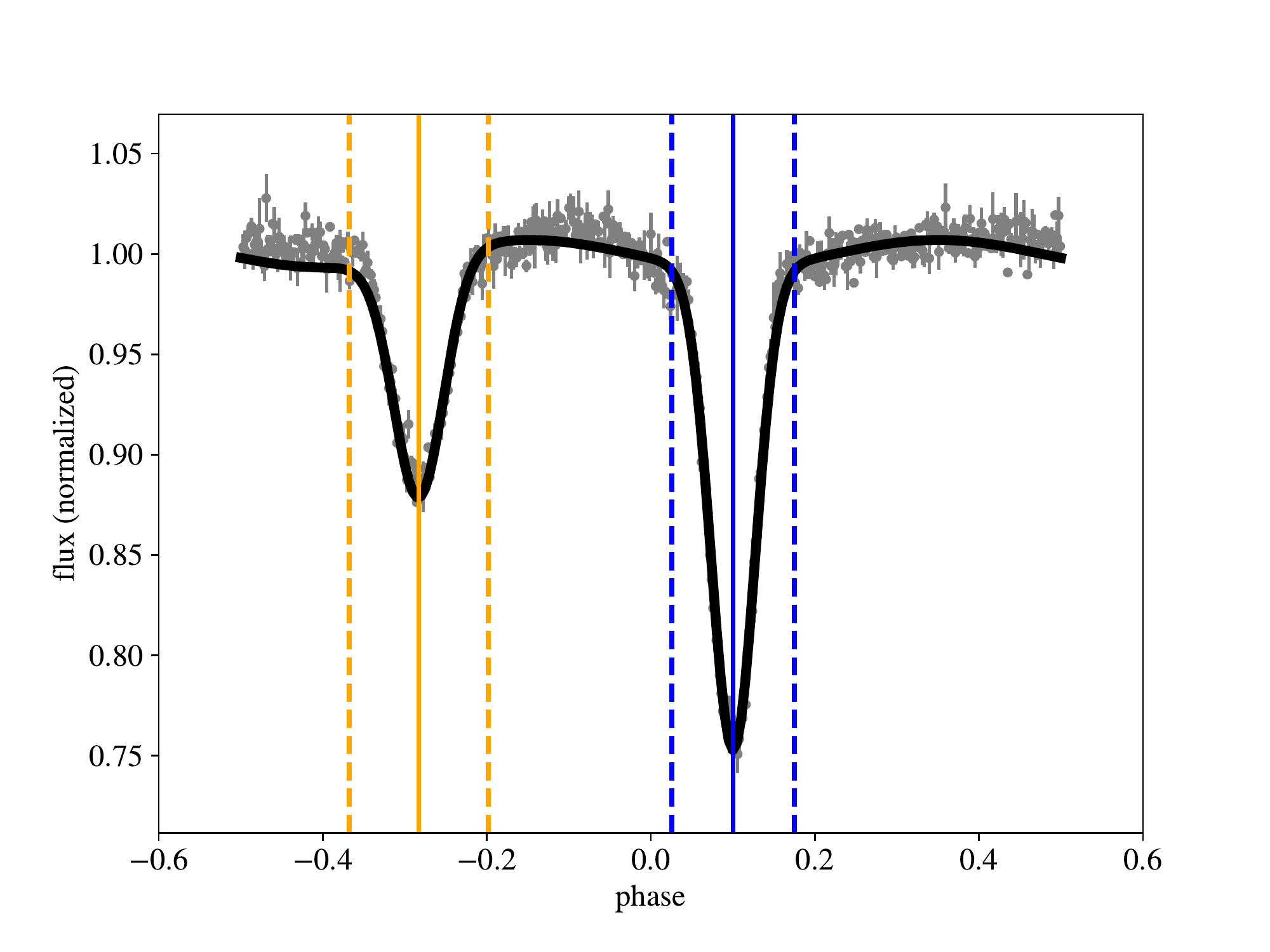}
    \caption{Best of seven trial two-Gaussian models used to determine the phases of eclipse minima, ingress, and egress (blue and orange vertical lines representing the primary and secondary eclipses, respectively) as well as the input binned synthetic observations.  These are then used to estimate values for eccentricity ($e$), argument of periastron ($\omega_0$), and the time of superior conjunction ($t_{0,\mathrm{supconj}}$).}
    \label{fig:lc_geometry}
\end{figure}

The two-Gaussian model is a fast, analytical model that fits a composite function of a constant, cosine term, two Gaussians, or any combination of these to the observed data, resulting in seven different models. The simplest model consists only of a constant term ($C$), while the most complex one of a constant, two Gaussian functions and a cosine term. Each Gaussian is characterized by its central position ($\mu$), amplitude ($A$), and root mean square (RMS) width ($\sigma$). The cosine term is only characterized by its amplitude ($A_{\mathrm{ell}}$) as the model assumes its period is equal to half the light curve period (0.5 in the case of a phase-folded light curve) and its trough coincides with either the position of the primary or secondary eclipse.  Figure \ref{fig:lc_geometry} shows an example case with the best fit model and the resulting eclipse phases.

The two-Gaussian model is sensitive to the initial values of the model parameters. Thus, to ensure its convergence, PHOEBE first estimates the eclipse positions, widths and depths using a simple algorithm that searches for the minimum of the light curve and isolates data in its vicinity that cross the median flux in phase ($\varphi$) space. All seven models are then fitted with a least-squares optimizer and the best fit is chosen as the one with the highest Bayesian Information Criterion \citep[BIC;][]{bic1978}. We determine the eclipse parameters following the prescriptions in \citet{mowlavi2017}, where the positions of the eclipses coincide with the central positions of the Gaussians ($\varphi_i = \mu_i$), the widths of the eclipses are a factor of the Gaussian RMS widths ($w_i = 5.6 \sigma_i$) and the depths are computed as the constant term minus the flux at eclipse positions ($d_i = C - \mathrm{flux}(\varphi_i)$).

The estimator computes the time of superior conjunction, eccentricity and argument of periastron as:

\begin{description}
    \item[time of superior conjunction]
    \begin{equation}
        t_{0, \mathrm{supconj}} = t_{0, \mathrm{supconj}, \mathrm{orig}} + \varphi_1 P_\mathrm{orb}
    \end{equation}
    \item[eccentricity]
    \begin{equation}
    e = \left[\sin^2\left(\frac{\psi-\pi}{2}\right) + \left(\frac{w_2-w_1}{w_2+w_1}\right)^2\cos^2\left(\frac{\psi-\pi}{2}\right)\right]^{1/2}
    \end{equation}
    \item[argument of periastron]
    \begin{eqnarray}
        \omega_1 & = & \arcsin\left(\frac{1}{e}\frac{w_2-w_1}{w_2+w_1}\right) \\
        \omega_2 & = & \arccos\left(\frac{\sqrt{1-e^2}}{2e\tan(\psi-\pi)}\right) \\
         \omega & = & \left\{
          \begin{array}{@{}cc@{}}
            \omega_2, & \text{if}\ \omega_1 >= 0 \\
            2\pi - \omega_2, & \text{if}\ \omega_1 < 0
          \end{array}\right.
    \end{eqnarray}

where $\psi = \pi + 2\arctan\frac{e\cos\omega}{\sqrt{1-e^2}}$ and is computed iteratively through solving $2\pi\Delta\Phi = \psi - \sin\psi$, where $\Delta\Phi$ is the separation between the two eclipses in phase space.
\end{description}

In addition, the phases of ingress and egress and eclipse widths and depths are computed and exposed to the user.  These are then used to also propose an optional phase-mask based on the eclipse edges and a padding of 30\% the eclipse width (see Section \ref{sec:solver_times} and Figure \ref{fig:lc_geometry_mask}).

\subsection{Artificial Neural Network (\texttt{\sc ebai})}\label{sec:ebai}

\texttt{\sc ebai} is a back-propagating artificial neural network specifically implemented to recover fundamental parameters from phased light curves of eclipsing binary systems \citep{prsa2008, 2019ascl.soft08018P}.  For the implementation within PHOEBE 2.3, we retrain the network on detached systems using the two-Gaussian model (see Section \ref{sec:lc_geometry}), using 201 input units, 40 hidden units and 5 output units, learning rate parameter $\eta=0.2$, 33,235 exemplars and $4 \times 10^6$ iterations.

\begin{figure}
    \centering
    \includegraphics[width=0.5\textwidth]{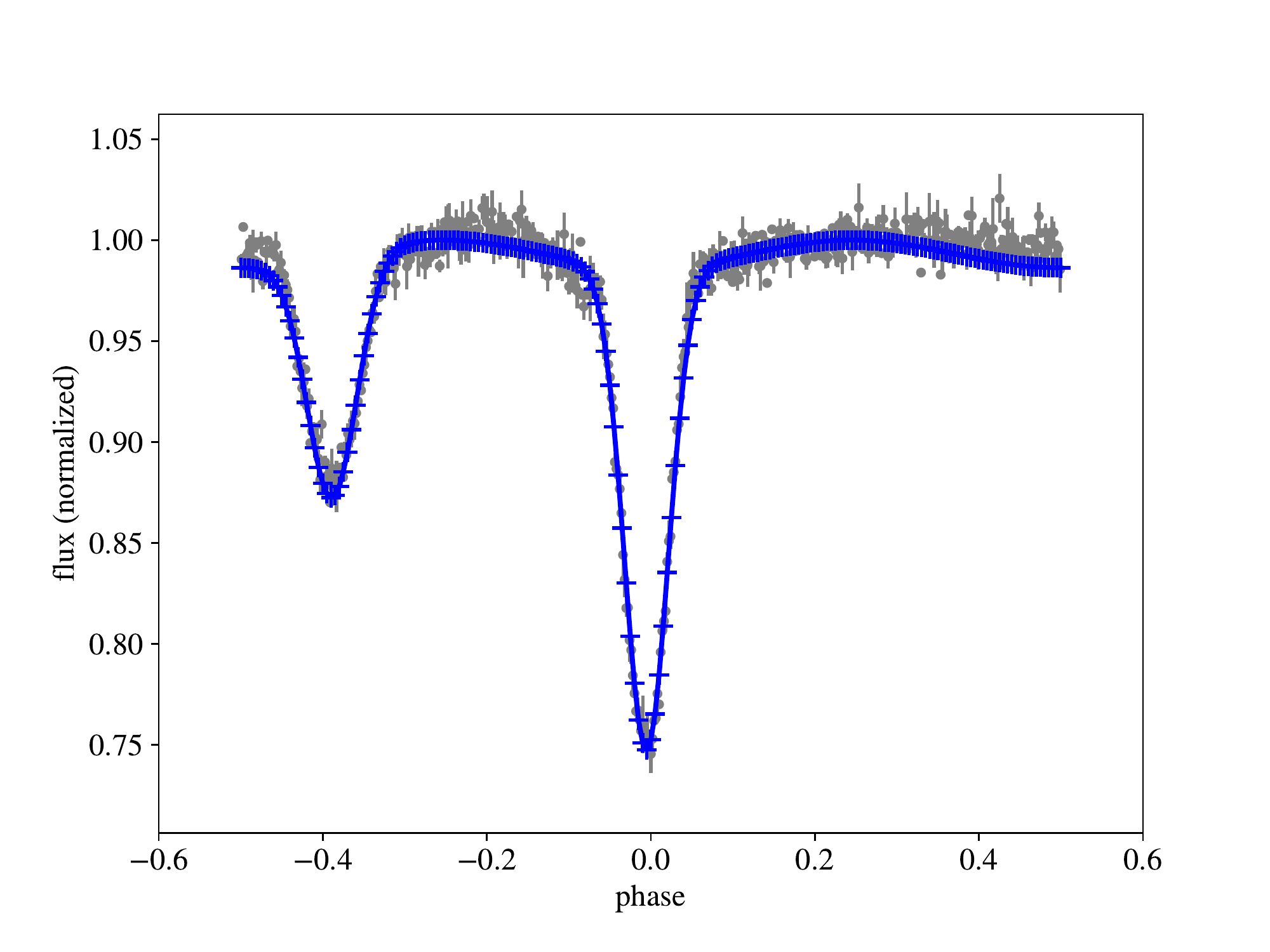} \\
    \caption{Analytic two-Gaussian representation of the automatically binned phased light curve sent to ebai to estimate $t_{0, \mathrm{supconj}}$, $T_\mathrm{eff, ratio}$, $R_\mathrm{equiv, fractional~sum}$, $e \sin \omega_0$, $e \cos \omega_0$, and $i$.}
    \label{fig:ebai}
\end{figure}

To use \texttt{\sc{ebai}} as an estimator, the user can pass any number of light curve data sets, which are normalized and fitted with a two-Gaussian model, automatically introducing a phase-shift if necessary to ensure the eclipse is at phase zero as required by the training set.  This analytic representation is then normalized to unity, sampled at the same 201 phase-points as the training set, and is passed to the two matrix transformations determined from the \texttt{\sc ebai} training (Figure \ref{fig:ebai}).  The results directly from \texttt{\sc ebai} are then converted as necessary to the proposed values for: $T_\mathrm{eff, 2} / T_\mathrm{eff, 1}$, $(R_\mathrm{equiv, 1} + R_\mathrm{equiv, 2}) / a_\textrm{orb}$, $e \sin \omega_0$, $e \cos \omega_0$, and $i$, in addition to $t_{0, \mathrm{supconj}}$ computed from the applied phase-shift.

\section{The Merit Function}\label{sec:merit_function}

Any optimization (see Section \ref{sec:optimizers}) or sampling (Section \ref{sec:samplers}) that requires the forward-model to be computed, requires a definition of the merit function to compare the synthetic model to the observations.  The components of the merit function (lnprobability) and terms that have been adopted within PHOEBE are explained below:

\begin{description}
    \item[lnpriors] the log-priors term is defined as the log probability of drawing the current face values, $p$, from the provided prior probability distribution functions, $\pi$.  When priors are not provided or applicable, this term will be constant and so will not contribute to the optimization/sampling.  Priors are discussed in more detail in Section \ref{sec:priors}.
    
    \begin{equation}\label{eq:lnpriors}
        \mathrm{lnpriors} = \sum_\mathrm{priors} \mathrm{ln}( \mathrm{P}(p \mid \Pi))
    \end{equation}
    
    \item[residuals] observations ($y_{o}$) $-$ synthetic model ($y_{m}$), whether those are fluxes or radial velocities, after any phase masking (see Section \ref{sec:solver_times}).  By default, the synthetic model is computed at the exact times of the (masked) observations.  However, the user can also compute the model at a list of custom times.  In that case, the synthetic model is linearly interpolated to the times (or phases) of the observations and then residuals are determined.  It is the responsibility of the user to ensure the model is sufficiently sampled so that linear interpolation does not cause any issues. See Section \ref{sec:forward_model} for details about the forward-model itself, times of the synthetic model, phase-masking, and interpolation.
    
    \begin{equation}
        \mathrm{residuals} = y_{o}(t_\mathrm{masked}) - y_{m}(t_\mathrm{masked})
    \end{equation}
    
    \item[chi2 ($\chi^2$)] $\chi^2$ is defined here to be the sum of squares of the residuals of all data points across all enabled ``datasets'' over the squares of the provided per point uncertainties (if provided - although note that some algorithms require uncertainties to be provided) plus an additional term to handle uncertainty underestimation from the provided observation uncertainties, $\sigma_\mathrm{o}$.  This uncertainty underestimation ($\sigma_\mathrm{lnf}$) term can be optimized or sampled in the case where the provided uncertainties are believed to be underestimated.  By default, $\sigma_\mathrm{lnf} = - \inf$, which cancels the additional term in the denominator.  See Section \ref{sec:noise_nuisance} for more about this ``noise-nuisance'' parameter.
    
    \begin{equation}\label{eq:chi2}
        \chi^2 = \sum_\mathrm{datasets} \frac{(y_{o} - y_{m})^2}{\sigma^2} + \mathrm{ln}(\sigma^2),
    \end{equation}
    where
    \begin{equation}\label{eq:chi2sigma}
        \sigma^2 = \sigma_\mathrm{o}^2 + y_{m}^2  \mathrm{e}^{(2  \sigma_\mathrm{lnf})}
    \end{equation}

    \item[lnlikelihood] the log-likelihood is, by definition:
    
    \begin{equation}
        \mathrm{lnlikelihood} = - 0.5 \chi^2
    \end{equation}
    
    \item[lnprobability] the log-probability is the default merit function used within PHOEBE (note that, technically, the negative log-probability is used for optimizers which minimize a merit function as the optimal model is found by maximizing the log-probability).  When priors are not provided or applicable, this essentially becomes analogous to $\chi^2$.  The log-probability is defined as the sum of the log-priors and the log-likelihood:
    
    \begin{equation}\label{eq:lnprobability}
        \mathrm{lnprobability} = \mathrm{lnpriors} + \mathrm{lnlikelihood}
    \end{equation}
    
    \item[lnposteriors] the log-posteriors, similar to the log-priors, describe the probability of drawing the current face values, $p$, given the resulting posterior probability distribution functions.  These posteriors (see Section \ref{sec:posteriors}) are often determined from sampling the log-probability with an algorithm such as MCMC (see Section \ref{sec:mcmc}).
    
    \begin{equation}
        \mathrm{lnposteriors} = \sum_\mathrm{posteriors} \mathrm{ln}( \mathrm{P}(p \mid \mathrm{posterior~pdf}))
    \end{equation}
    
\end{description}

Unless overriding with a custom merit function, all optimizers and samplers within PHOEBE use the log-probability as defined above.  Additionally, if the user would like to implement their own optimizer or sampler outside of PHOEBE, all of the quantities above are exposed to the user for any given forward-model.

\subsection{Parameterization}\label{sec:parameterization}

Depending on the available observations (single light curve, multiple light curves, single-lined radial velocities, double-lined radial velocities, etc.), previously known information, and the orthogonality (or lack thereof) of the parameter-space, it is advantageous to be able to choose a specific parameterization.  For example, due to the parameterization of the Roche model, PHOEBE 1 (based on Wilson-Devinney) parameterizes the orbit by the orbital period ($P_\mathrm{orb}$), mass-ratio ($q$), and semi-major axis ($a_\mathrm{orb}$) (along with eccentricity ($e$) and orientation parameters such as inclination, and argument of periastron).  The individual stellar masses ($M_1$ and $M_2$) can be computed from these parameters and Kepler's third law, but cannot directly be fixed or adjusted.

PHOEBE now provides a flexible framework for scenarios like these.  For example, by default, PHOEBE adopts a similar parameterization to the legacy version but also exposes read-only parameters for the masses that are derived by Kepler's third law:

\begin{equation}
    M_1 = \frac{4 \pi a^3}{G P^2} \frac{1}{1+q}, \quad
    M_2 = \frac{4 \pi a^3}{G P^2} (1+q)
\end{equation}

This system of five parameters ($M_1$, $M_2$, $P$, $a$, $q \equiv M_2/M_1$) can be manipulated from the python interface such that any two of the five parameters are ``constrained'' as read-only, based on Kepler's third law.  Setting or fitting the individual masses is then possible by choosing two other parameters to be read-only instead. 

There are many other similar parameterization choices in a binary model in which one parameterization may be preferred over another due to the orthongonality within the local parameter space or due to the availability of known information in advance. Table \ref{table:constraints} shows a list of the applicable constraints included in the 2.3 release of PHOEBE.

\begin{center}
\begin{table}
\begin{tabular}{ |c|c||c| }
\hline
  Default Free Parameter(s) & Other Inputs  & Default Read-Only Parameter(s) \\
 \hline \hline
 $P_\mathrm{orb}$, $q$, $a_\mathrm{orb}$ & & $M_1$, $M_2$ \\
 \hline
 $P_\mathrm{orb}$ & & $f_\mathrm{orb}$ (frequency) \\
 \hline
 $e$, $\omega_0$ & & $e \sin \omega_0$, $e \cos \omega_0$ \\
 \hline
 $a_\mathrm{orb}$, $i_\mathrm{orb}$ & &  $a_\mathrm{orb} \sin i_\mathrm{orb}$ \\
 \hline
 $t_{0,\mathrm{supconj}}$ & $P\mathrm{orb}$, $\dot{P}_\mathrm{orb}$ $e$, $\omega_0$, $\dot{\omega}$, $t_0$ & $t_{0,\mathrm{perpass}}$, $t_{0,\mathrm{ref}}$ \\
 \hline
 $t_{0,\mathrm{perpass}}$ & $P_\mathrm{orb}$, $\dot{P}_\mathrm{orb}$, $t_0$ & $M$ (mean anomaly) \\
 \hline
 $F$ (synchronicity) & $P_\mathrm{orb}$ & $P_\mathrm{rot}$ \\
 \hline
 $P_\mathrm{rot}$ & &  $f_\mathrm{rot}$ (frequency) \\
 \hline
 $M$, $R_\mathrm{equiv}$ & & logg \\
 \hline
 $a_\mathrm{orb}$ & $q$ & $a_\mathrm{comp}$ \\
 \hline
  $a_\mathrm{orb}$ & $i_\mathrm{orb}$, $q$ &  $a_\mathrm{comp} \sin i_\mathrm{orb}$ \\
 \hline
 pitch, $i_\mathrm{orb}$ & & $i_\mathrm{comp}$ \\
 \hline
 yaw, $\Omega_\mathrm{orb}$ (asc.~node) & & $\Omega_\mathrm{comp}$\\
 \hline
 $A_v$, $R_v$ & & $E(B-V)$\\
 \hline
 contacts: $R_\mathrm{equiv}$ & $q$ & $FF$ (fillout factor), $\Omega$ (equipotential) \\
 \hline 
 \multicolumn{3}{c}{Optional Constraints} \\
 \hline
 $T_{\mathrm{eff},1}$, $T_{\mathrm{eff},2}$ & & $T_\mathrm{eff, 2} / T_\mathrm{eff, 1}$ \\
 \hline
 $R_{\mathrm{equiv},1}$, $R_{\mathrm{equiv},2}$, $a_\mathrm{orb}$ &  & $R_\mathrm{equiv,2} / R_\mathrm{equiv,1}$, $(R_\mathrm{equiv, 1} + R_\mathrm{equiv, 2}) / a_\textrm{orb}$\\
 \hline
 semidetached: $R_\mathrm{equiv}$ & & $R_\mathrm{equiv, crit}$ \\
 \hline
 $\pi$ (parallax) & & $d$ (distance) \\
 \hline 
\end{tabular}
\caption{Built-in constraints in PHOEBE (the top of the table before the break are included in systems by default whereas the bottom can optionally be added).  The parameters in the right column are, by default, read-only and derived from the parameters in the two left columns.  The read-only status of any parameter in the right column can be replaced with one in the left column (as long as any given parameter is ``constrained'' by a single expression), allowing for a large combination of parameterization choices.  Parameters in the middle column are included in the expression itself, but are not currently available to be chosen as the read-only parameter.}
\label{table:constraints}
\end{table}
\end{center}

\subsection{Priors}\label{sec:priors}

PHOEBE allows creating and attaching distributions (as \texttt{\sc distl}\footnote{https://github.com/kecnry/distl} distribution objects) to any parameter represented by a float-value in the system.  \texttt{\sc distl} is a Python package which handles defining, converting, storing, and manipulating various probability distributions.  The version currently packaged with PHOEBE supports the following univariate distribution types: uniform (boxcar), Gaussian, delta, histogram, and samples (which generates a Kernel Density Estimation -- KDE -- under-the-hood around an input set of samples to allow drawing from the same distribution as the samples while not drawing from the set directly).  Additionally, the following multivariate distribution types are supported: multivariate gaussians, multivariate histograms, multivariate samples which each contain the known covariances between parameters.

\begin{figure}
    \subfloat{%
      \includegraphics[width=0.33\columnwidth]{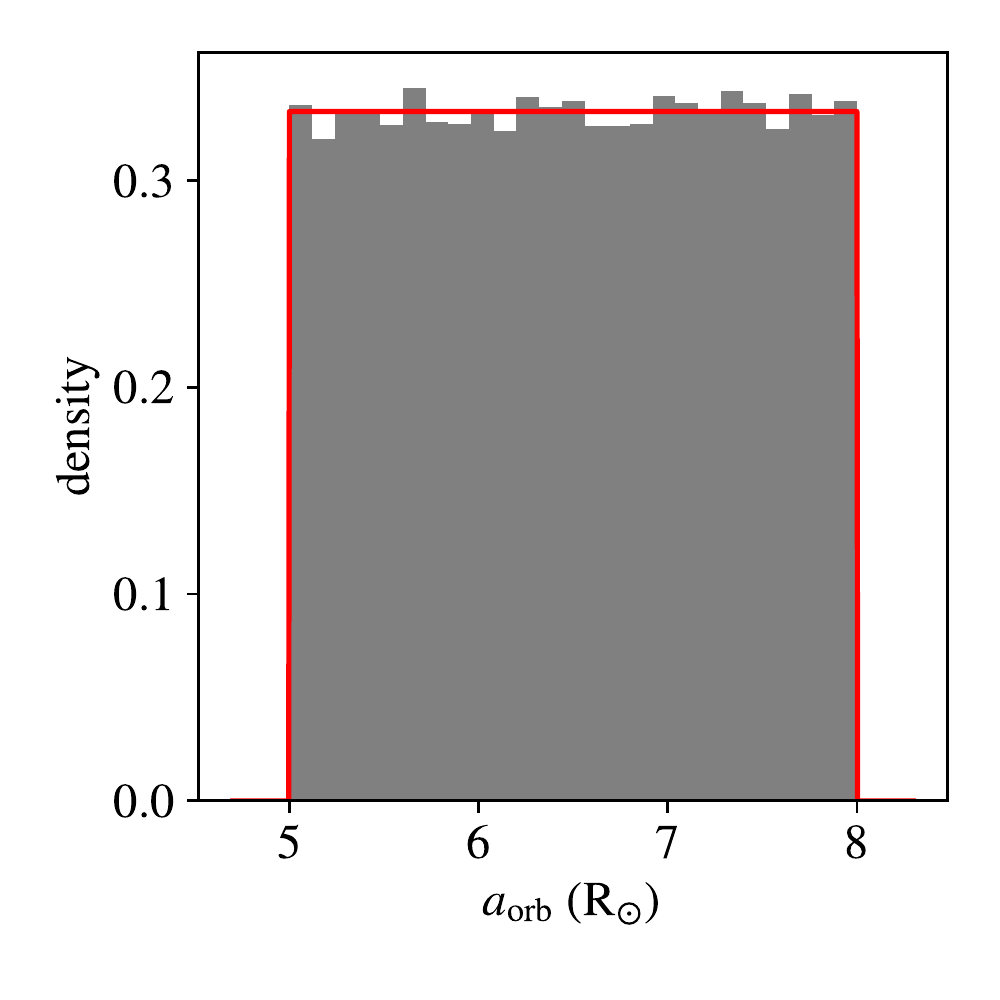}
    }
    \subfloat{%
      \includegraphics[width=0.33\columnwidth]{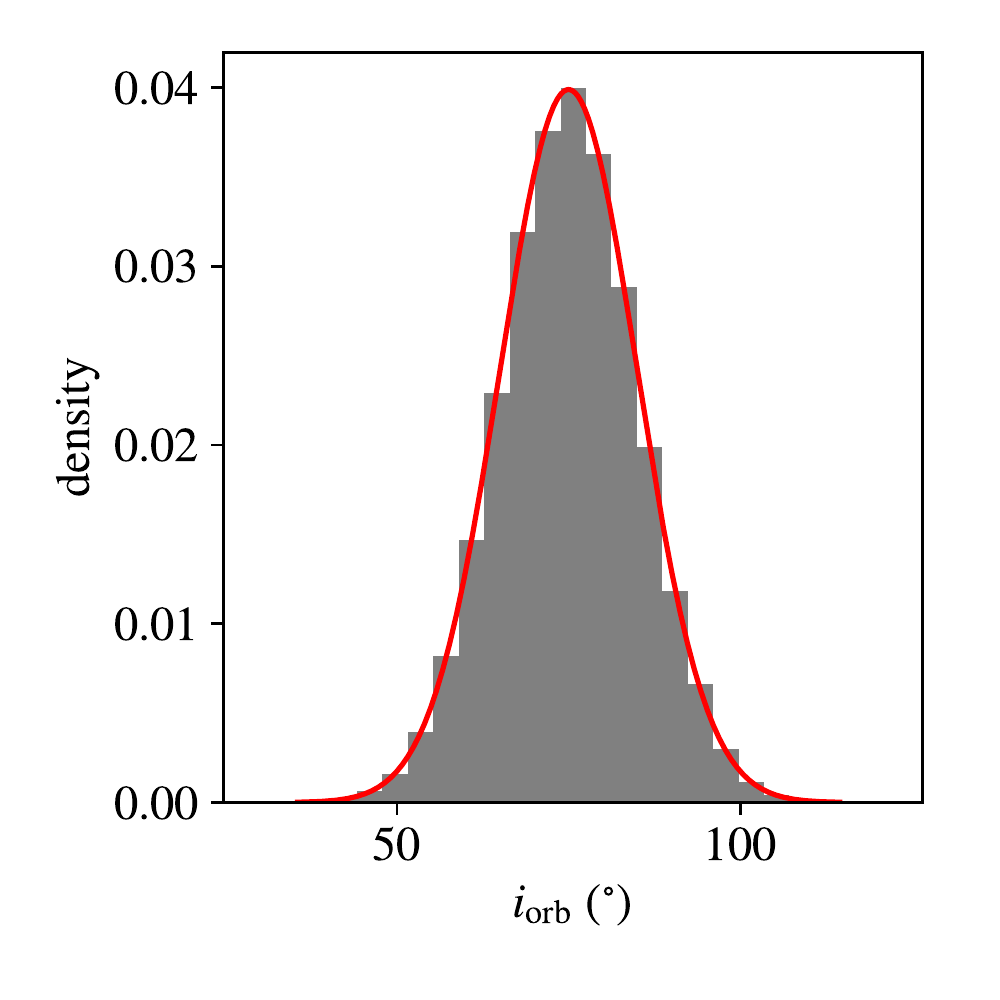}
    }
    \subfloat{%
      \includegraphics[width=0.33\columnwidth]{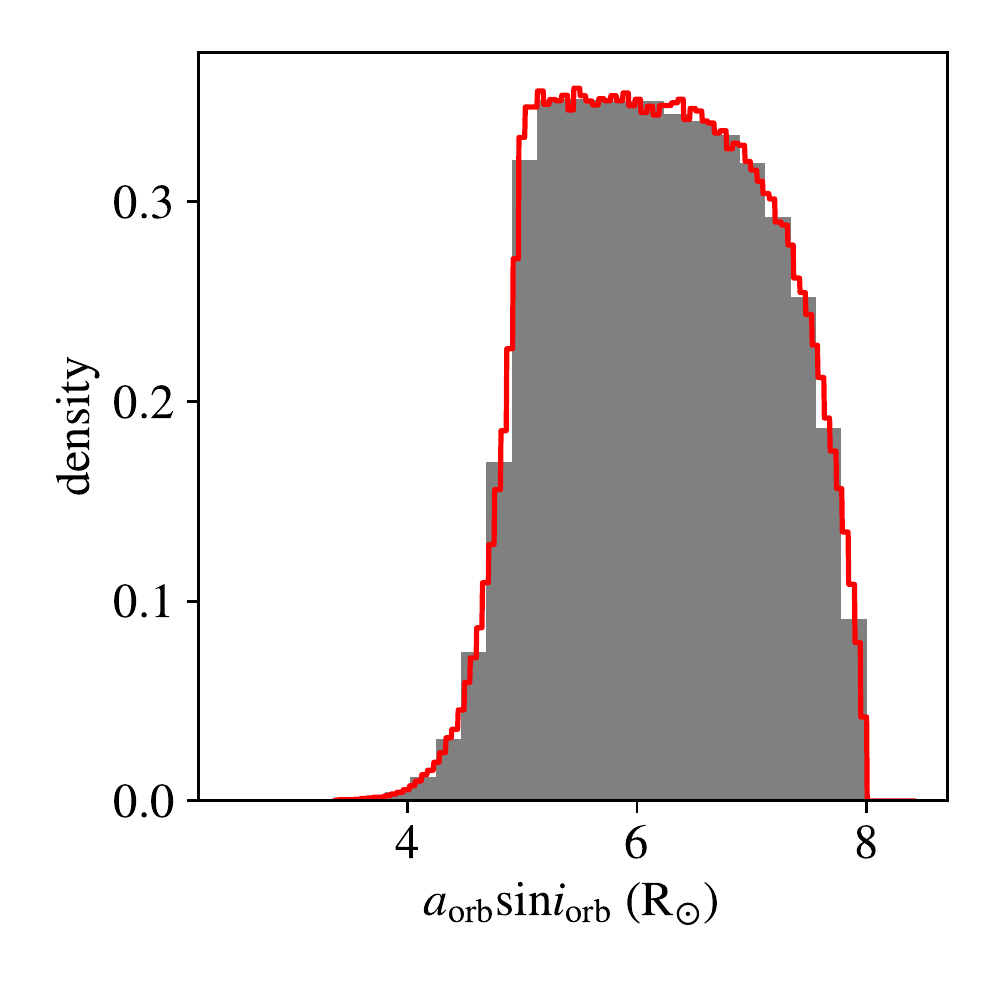}
    }
    \caption{A uniform distribution placed on the semi-major axis (left) and a gaussian distribution placed on the inclination (middle).  These distributions can then be propagated through the logic discussed in Section \ref{sec:parameterization} to get the resulting distribution on $a \sin i$ (right).}
    \label{fig:priors_constraints}
\end{figure}

Once defined and attached to a parameter, these distributions can then be plotted directly, or propagated through the constraint logic discussed in Section \ref{sec:parameterization} (see Figure \ref{fig:priors_constraints}).  Any distribution attached to a ``free'' parameter can have its value randomly drawn and set for the respective distribution.  Any distribution (whether attached to an adjustable or ``read-only'' parameter) can provide the probability of drawing the current face value of the parameter.  When passed to an optimization or sampling algorithm that supports priors, these probabilities are then included in the merit function (see Equations \ref{eq:lnpriors} and \ref{eq:lnprobability}).

This flexibility allows for numerous useful scenarios.  For example, it is possible to set a prior distribution on $a \sin i$ (e.g., known from the literature) while marginalizing over a defined range in both $a$ and $i$.  This also allows for setting a prior on any observationally-constrained parameter that is not directly a dataset observable, including $\log g$ or mass from spectra reduction, for instance. As will be discussed in Section \ref{sec:posteriors} it is also possible to sample with one parameterization ($e$ and $\omega_0$, for example) but then expose the posteriors and uncertainties in another ($e \sin \omega_0$ and $e \cos \omega_0$).

\subsection{Angle-Wrapping}\label{sec:angle_wrapping}

All parameters and distributions that correspond to angles (argument of periastron, longitude of ascending node, etc.) include automatic angle-wrapping (with the exception of inclination which is limited to the range between $0$ and $\pi$ to avoid ambiguity with the longitude of ascending node). For example, if setting a value, either manually or within an optimizer or sampler, outside the $[0, 2 \pi]$ range, the value will automatically be mapped back onto the range between $[0, 2 \pi]$.  This is particularly useful if the true value is near the wrapping point -- as the optimizer or sampler can explore the parameter space continuously.

This can cause some issues within optimizing or sampling when the merit function is insensitive to one of the parameters.  For example, for a circular system, the argument of periastron has no effect on the observables.  If allowed to wrap continuously, the sampler will continue to randomly walk indefinitely.  To avoid this, the per-wrapped angles are limited to within $\pi$ from the central value of the initializing distribution (see Section \ref{sec:samplers}), essentially placing an uninformative prior and preventing the sampler from wandering across the wrapped space.

\subsection{Forward-Model}\label{sec:forward_model}

The ``forward-model'' consists of the synthetic model for a fixed set of input parameters, which is then compared to the observations through the merit function as described in detail in Section \ref{sec:merit_function}.  

\subsubsection{Phase-Masking, Exposure Times, Undersampling, and Interpolation}\label{sec:solver_times}

\begin{figure}
    \centering
    \includegraphics[width=0.5\textwidth]{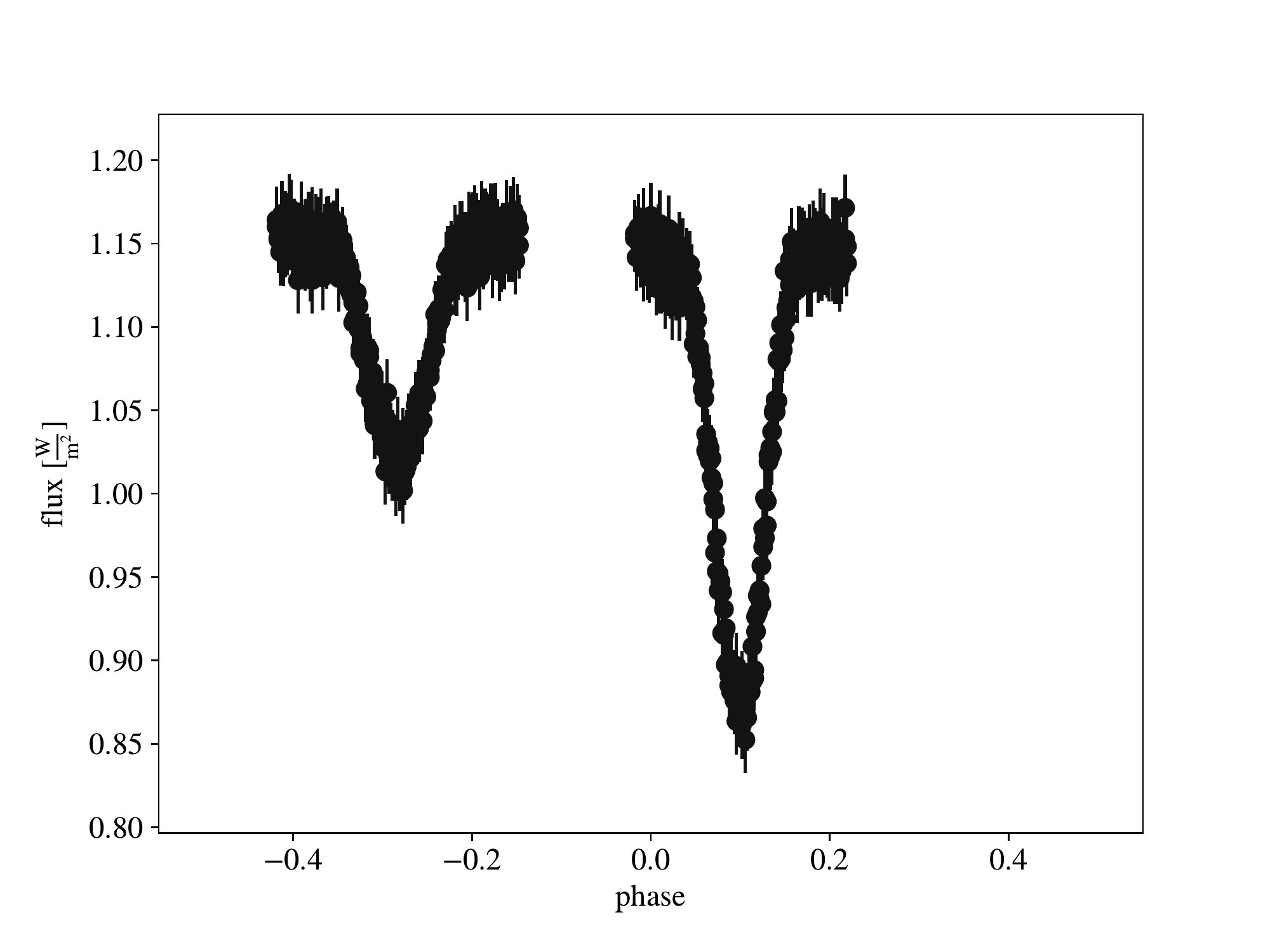}
    \caption{The same light curve data set as in Figures \ref{fig:lc_geometry} and \ref{fig:ebai} masked to the phases of ingress and egress (plus an additional 30\% padding) as proposed by the light curve geometry estimator (see Section \ref{sec:lc_geometry}).}
    \label{fig:lc_geometry_mask}
\end{figure}

By default, PHOEBE will compute the forward-model at the timestamps of the observations.  For convenience purposes, PHOEBE also supports excluding specific phases of the data to improve computational efficiency.  For example, during initial exploration and optimizations of basic geometric and orbital parameters, it can be useful to exclude the out-of-eclipse section of light curves (see Figure \ref{fig:lc_geometry_mask}).  These data should then be re-introduced into the merit function for further optimization and determination of posteriors from all available observations.  To allow for this, PHOEBE implements a very basic ``phase-masking'' which can be enabled and disabled.  Note however that any outlier-removal or more complex masking (in time-space, for example), will currently need to be done outside of PHOEBE.  When phase-masking is being used within optimizing or sampling, PHOEBE will exclude those times from both the forward-model and the residuals in the merit function.

Additionally, it is often useful to ``downsample'' the observations to save on computation time or, when not optimizing or sampling, to ``oversample'' for plotting purposes or to predict how the model behaves outside of the limits of observations.  To accomplish all of these use cases, PHOEBE allows overriding the computation times of each data set to a custom array of times.  When optimizing or sampling, PHOEBE will, by default, automatically take the shorter of these two arrays (the masked observation times or the requested computation times).  Within the merit function, the synthetic model is then interpolated onto the times of the masked observations.  In cases where the system is not time-dependent, PHOEBE will allow for ``extrapolating'' outside the time-bounds of the synthetic model by interpolating in phase-space, but will not allow doing so when the system is time-dependent (for example: if any time-derivative is non-zero, including the presence of apsidal motion, or at least one of the components has spots and is asynchronously rotating).  

Lastly, PHOEBE supports oversampling the model over the exposure time for each requested synthetic time.  The resulting synthetic flux at the mid-time is then exposed as the mean of the oversampled fluxes.  Long-cadence Kepler data, for example, has a 29.4 minute exposure time which results in phase-smoothing of the photometry.  Setting the appropriate exposure time and enabling oversampling allows the model to reproduce this effect.

\subsubsection{Limb-Darkening}\label{sec:ld_mode}

PHOEBE 2.0 introduced automatic interpolated limb-darkening by default \citep[see][Section 5.2.3]{prsa2016}, in addition to manual coefficients for linear, logarithmic, quadratic, square-root, and power laws.  PHOEBE 2.2 \citep{jones2020} then included the ability to automatically query coefficients for any of the built-in laws, on a per-surface-element basis.

Optimizing or sampling in PHOEBE 2.3 supports both interpolated and ``lookup'' limb-darkening, but does not currently support fitting manually set coefficients.  Interpolated limb-darkening and per-element lookup is only supported within the PHOEBE backend itself -- for other backends (see Section \ref{sec:alternate_backends}), the coefficients are queried from the atmosphere tables in PHOEBE based on the mean stellar parameters.

\subsubsection{Alternate Computation Backends}\label{sec:alternate_backends}

PHOEBE has been designed to be as robust and accurate as possible \citep[for specific implementation details, see ][]{prsa2016, horvat2018, jones2020} -- but this does come at the cost of computational efficiency.  Any inefficiency is magnified when a large number of forward-model instances needs to be generated by an optimizer or sampler.  Depending on the specific system being studied, substantial time can be saved by ``disabling'' certain effects and making simplifying assumptions.  PHOEBE does support some of these assumptions -- irradiation is expensive and can be disabled when negligible or spherical stars can replace Roche-distorted stars, for instance.  Other codes may be better suited or more optimized for a given system.  In some cases, higher-order effects that require the use of PHOEBE can be ignored while searching the whole parameter space and optimizing orbital parameters, for example, but may still be necessary to include when attempting to get precise estimates for the final reported values.

\begin{figure}
    \subfloat{%
      \includegraphics[width=0.5\columnwidth]{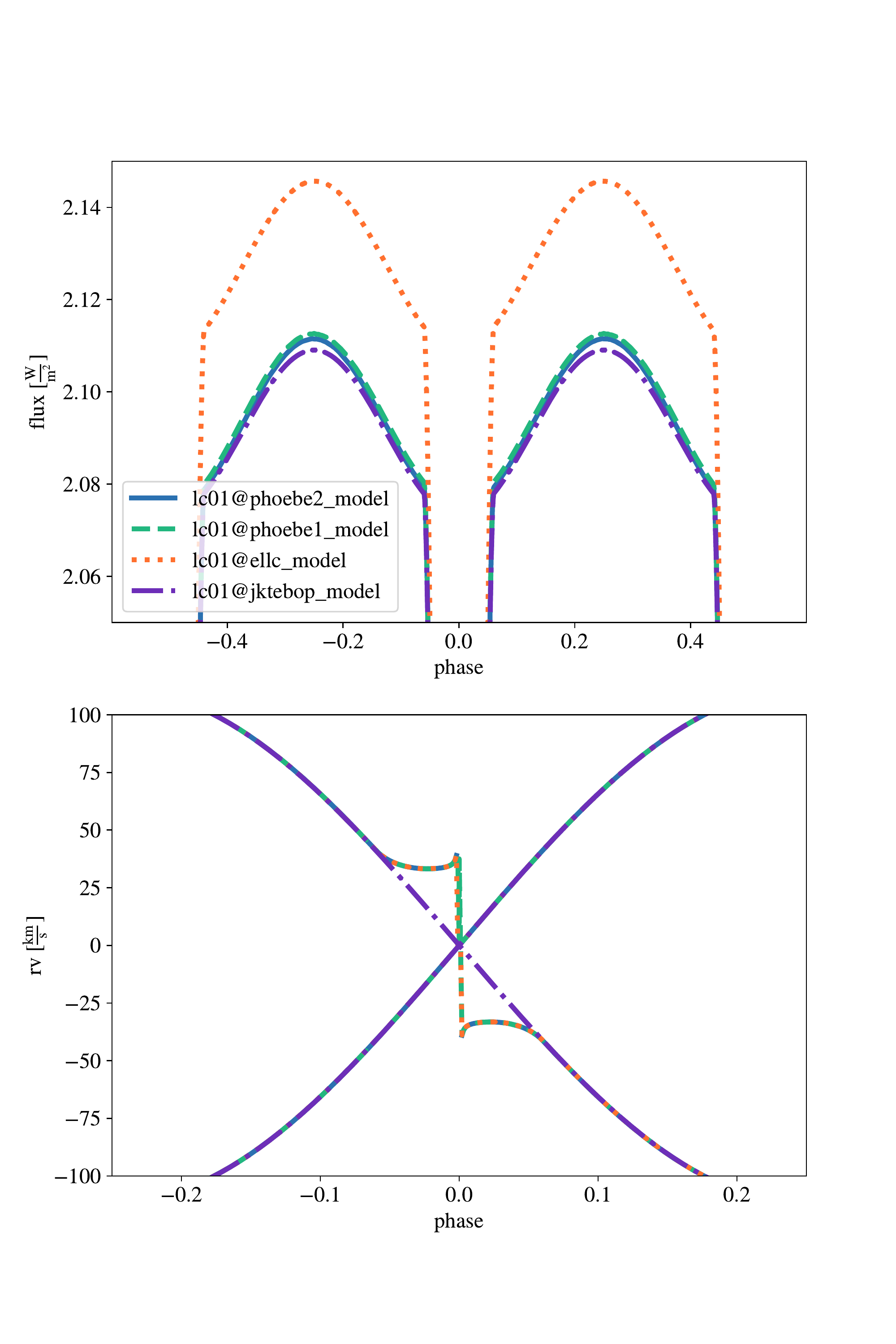}
    }
    \subfloat{%
      \includegraphics[width=0.5\columnwidth]{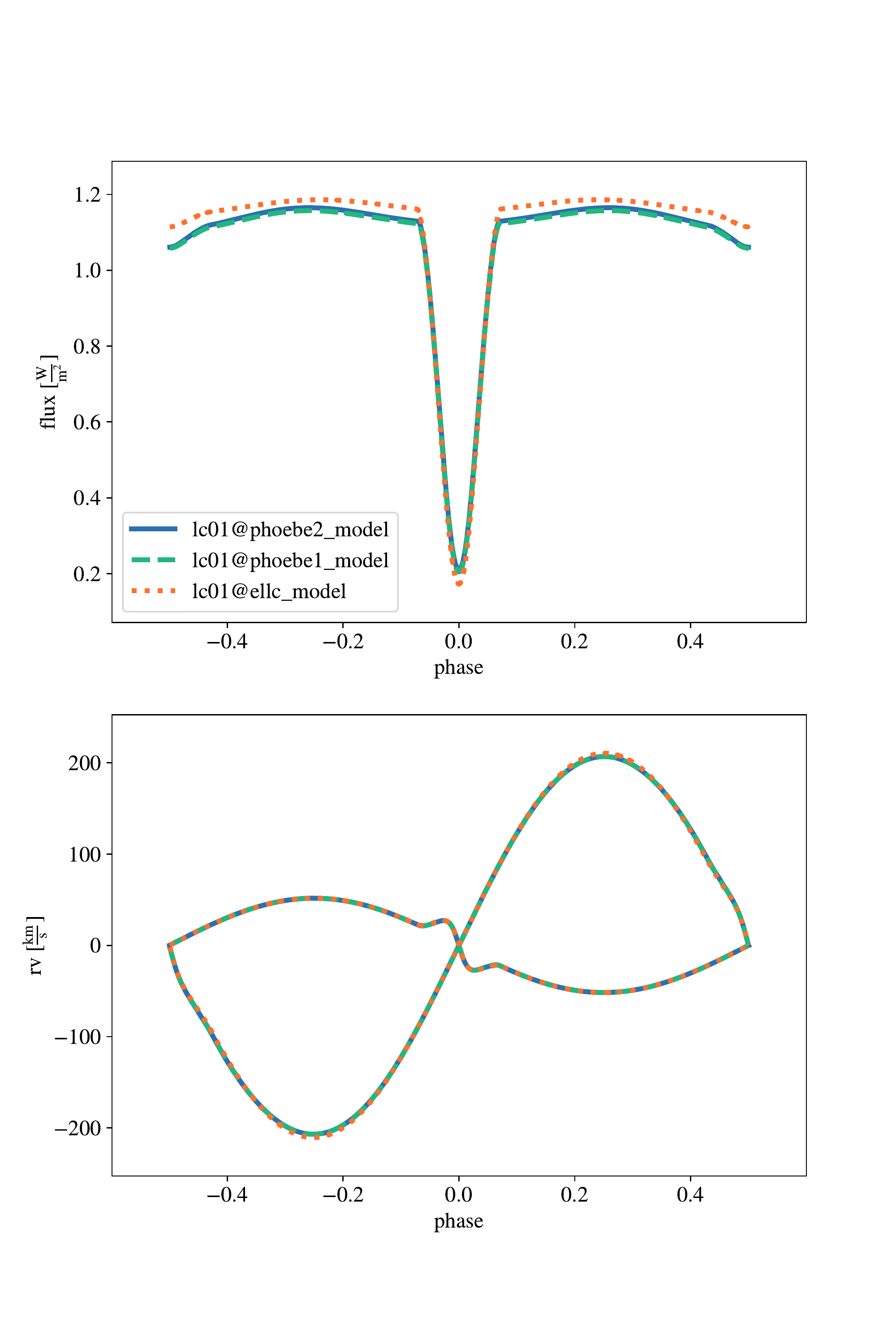}
    }
    \caption{Comparison of several supported backends run through PHOEBE on the same system (detached system on the left and semi-detached system on the right).  Although not identical, this is useful to compare the assumptions between different models and to conveniently take advantage of models that are most efficient for a particular science-case or system.  Slight vertical scaling differences are likely due to the assumptions in each code as well as the assumptions in the flux-scaling described in Section \ref{sec:flux_scaling}.}
    \label{fig:backends_compare}
\end{figure}

For this reason, and to enable comparing codes against each other, PHOEBE includes wrappers to the forward-model component of several other public codes, including: PHOEBE legacy \citep[based on Wilson-Devinney;][]{prsa2005, wilson1971, wilson1979, wilson2008, wilson2014}, \texttt{\sc ellc} \citep{maxted2016}, and \texttt{\sc jktebop} \citep{southworth2004, southworth2007, southworth2009, southworth2011}.  These forward-models are designed to be ``drop-in'' replacements to the forward-model provided by PHOEBE, with a minimal number of additional options that are specific to that code.  Figure \ref{fig:backends_compare} shows a comparison of several forward-models computed through PHOEBE on the same set of parameters.  There are some notable discrepancies between the models caused by the different assumptions in each code and limitations on the ability to translate parameterizations and outputs of each code.  However, this interface provides the ability to conduct these comparisons and choose the most appropriate model for any given system.  Table \ref{table:backends} shows an overview of the implemented features, limitations, and assumptions of each of the backends currently available.  

These wrappers translate from the parameterization used within PHOEBE to the parameterization used within the requested forward-model, call the external code, and translate the results into the correct units used by PHOEBE.  It is important to note that this translation may not always be exact and does add some time-cost -- a native optimization and sampling wrapped around the individual codes will always perform faster.  However, regardless of the chosen forward-model, all backends can also benefit from some of the functionality built into the PHOEBE frontend, including: flexible parameterization via ``constraints'' (see Section \ref{sec:parameterization}), phase-masking, exposure times, and time-interpolation (see Section \ref{sec:solver_times}), limb-darkening coefficient lookups (see Section \ref{sec:ld_mode}), flux-scaling via passband luminosities and third light in light curves (some natively, see Section \ref{sec:flux_scaling}), systemic velocities and per-component radial velocity offset (see Section \ref{sec:rv_offset}), noise nuisance to handle observational uncertainty underestimation (see Section \ref{sec:noise_nuisance}), gaussian processes (see Section \ref{sec:gaussian_processes}), and all of the optimization and sampling methods implemented in PHOEBE.  If using any of these backends or disabling physical effects within the PHOEBE backend, it is always good practice to occasionally compare the resulting model with one computed without these assumptions.

Whenever publishing work that makes use of these alternate backends, we encourage citing the external code as appropriate.

\begin{center}
\begin{table}
\small
\begin{flushleft}
\begin{tabular}{ r||p{3cm}|p{3cm}|p{3cm}|p{3cm}|| }
  $ $ & PHOEBE 2 & PHOEBE 1 \newline(``legacy'') & ellc & jktebop \\
 \hline \hline
 Supported Versions & 2.3+ & 1.0 & 1.8.2+ & v40+ \\
 \hline
 LCs & yes & yes & yes & yes \\
 LCs (absolute fluxes) & yes & yes & re-scaled & re-scaled \\
 RVs (dynamical) & yes & yes & yes & yes \\
 RVs (with R-M) & yes & yes & yes (no irrad.) & no \\
 Spectral Line Profiles & forward-model & no & no & no \\
 Orbits & forward-model & no & no & no \\
 Access to Underlying Meshes & forward-model & forward-model & no & no \\
 \hline
 Detached Systems & yes & yes & yes & yes \\
 Semi-detached Systems & yes & yes & yes & no \\
 Contact Systems & yes & yes & no & no \\
 \hline
  (Native) Parameterization & period, \newline mass-ratio, \newline semi-major axis, \newline synchronicities, \newline equivalent radii, \newline Teffs, \newline eccentricity ($e$), \newline arg.~of per.~($\omega_0$) & period, \newline mass-ratio, \newline semi-major axis, \newline synchronicities, \newline equipotentials, \newline Teffs, \newline eccentricity ($e$), \newline arg.~of per.~($\omega_0$) & period, \newline mass-ratio, \newline \newline synchronicities, \newline fractional radii, \newline SB ratio, \newline $\sqrt{e} \cos \omega_0$, \newline $\sqrt{e} \sin \omega_0$ & period, \newline mass-ratio, \newline sum of fractional radii, \newline ratio of radii, \newline SB ratio, \newline $e \cos \omega_0$, \newline $e \sin \omega_0$ (or $e \& \omega_0$)\\
 \hline
 Surface Distortion & roche, \newline rotating star, \newline sphere & roche & roche, \newline sphere, \newline roche\_v, \newline poly1p5, \newline poly3p0, \newline love & biaxial-spheroid for ellipsoidal and irradiation contributions, sphere for eclipse shapes \\
 \hline
 Asynchronous Rotation & yes & yes & yes & no \\
 \hline
 Atmospheres & blackbody, \newline Castelli-Kurucz, \newline phoenix & blackbody (planckint), \newline Castelli-Kurucz (atmx) & none & none \\
 \hline
Number of Supported Passbands & $\sim 30$ & $\sim 30$ & 0 & 0 \\
 \hline
 Irradiation & Horvat/Lambert, \newline Wilson & Wilson & Lambert & biaxial-spheroid \\
 \hline
 Limb-Darkening & interpolated, lin, log, quad, sqrt, power & lin, log, sqrt & lin, log, quad, sqrt, power & lin, log, quad, sqrt \\
 \hline
 Gravity Brightening/Darkening & yes & yes & yes & yes \\
 \hline
 Spots & circular & circular & circular & no \\
 \hline
 R{\o}mer Delay & yes (optional) & no & yes & no \\
 \hline
 Apsidal Motion & yes & yes & yes & no \\
 \hline
 Period Time-Derivative & yes & yes & no & no \\
 \hline
 Spin-Orbit Misalignment & yes & no & limited & no \\
 \hline
 Extinction/Reddening & yes & no & no & no \\
 \hline \hline
\end{tabular}
\end{flushleft}
\caption{Comparison of available physics and features in the forward-model ``backends'' as wrapped by PHOEBE. Note that some additional capabilities may be available in the native versions of the codes but are excluded here if they are not easily mapped from PHOEBE's parameterization.}
\label{table:backends}
\end{table}
\end{center}

\subsubsection{Flux Scaling (Passband Luminosity, Distance, \& Third Light)}\label{sec:flux_scaling}

The surface fluxes (formally computed at a distance of 1\,m from the equivalent point source) calculated directly by the PHOEBE 2 backend are defined as \citep[cf.][]{prsa2018}: 

\begin{equation}\label{eq:F_backend_phoebe}
    F_{\mathrm{pb}|w} = \mathcal P_{\mathrm{int}|w} \sum_\mathrm{stars} S_\mathrm{rel}  \sum_\mathrm{elems} \langle I_{\mu, \mathrm{abs} |w} \rangle \mathscr{V} \mu \Delta A,
\end{equation}
where $|w$ denotes the flux weighting scheme (energy or photon counts), $\mathcal P_\mathrm{int|w}$ is the suitably weighted integral of the passband transmission function, ``stars'' are the components defined in the system, $S_\mathrm{rel}$ is the per-star scaling factor that translates between absolute and relative units, $\langle I_{\mu, \mathrm{abs}|w} \rangle$ are passband-averaged, per-surface-element specific emergent intensities interpolated from the model atmosphere tables, $\mathscr{V}$ is surface element visibility function, and $\mu \Delta A$ are projected surface areas of each element.
For components where a passband luminosity is provided by the user, $S_\mathrm{rel}$ is determined as:
\begin{equation}\label{eq:S_rel}
    S_\mathrm{rel} = \frac{L_{\mathrm{pb, rel}|w}}{L_{\mathrm{pb, abs}|w}}.
\end{equation}
Note that these absolute and relative luminosities are defined at time $t_0$ and exclude any extrinsic or aspect-dependent effects (distance, irradiation, spots, pulsations, boosting, etc).  Absolute luminosities are determined from absolute normal intensities from the tables:

\begin{equation}\label{eq:L_pb_abs_phoebe}
    L_\mathrm{pb, abs|w} = \pi \mathcal P_\mathrm{int|w} \sum_\mathrm{elems} \langle I_{\perp,\mathrm{abs}|w} \rangle \mathcal{D}_\mathrm{int} \Delta A,
\end{equation}
where $\langle I_{\perp, \mathrm{abs}|w} \rangle$ are the passband-averaged absolute emergent normal intensities interpolated from the model atmosphere tables, $\mathcal{D}_\mathrm{int}$ are the integrals of the limb darkening functions over all angles, and $\Delta A$ are the surface areas of each element \citep{prsa2018}.

Since version 2.2 \citep{jones2020}, PHOEBE has supported multiple modes for handling passband luminosities and scaling of light curve fluxes:

\begin{description}
    \item[absolute] intensities are used and integrated directly as retrieved from atmosphere and passband tables, resulting in absolute fluxes.  In this case, the scaling factor, $S_\mathrm{rel}$, in Equations (\ref{eq:F_backend_phoebe}) and (\ref{eq:S_rel}) is, by definition, unity.
    
    \item[dataset-scaled] if observations are provided, \textit{``dataset-scaled''} allows to automatically scale the resulting absolute fluxes to the data using a least squares algorithm.  This is particularly useful to optimize parameters for a normalized light curve without having to worry about correlations with the luminosity.  Although convenient, when determining uncertainties from sampling (Sections \ref{sec:samplers} and \ref{sec:posteriors}), it is advised to instead marginalize over the passband luminosities in order to avoid underestimated uncertainties.
    
    \item[component-coupled] a passband luminosity ($L_{\mathrm{pb, rel}|w}$) for one of the components is provided by the user, and intensities are internally scaled such that this prescribed intrinsic luminosity at time $t_0$ is achieved (Equation \ref{eq:S_rel}).  This same scaling factor is then used for both components in Equation (\ref{eq:F_backend_phoebe}).
    
    \item[dataset-coupled] if multi-band photometry is available, \textit{``dataset-coupled''} allows using the same scaling factor across multiple light curves to preserve any color information -- adopting $S_\mathrm{rel}$ for both components in one data set from those defined in another data set, and therefore allowing $L_{\mathrm{pb, rel}|w}$ to be computed from Equation (\ref{eq:S_rel}).

    \item[decoupled] decoupling allows for setting the desired luminosities of stars independently, ignoring any scaling information from the atmosphere and passband tables.  Here the scaling factors for each star in Equations (\ref{eq:F_backend_phoebe}) and (\ref{eq:S_rel}) are allowed to be independent of each other.
\end{description}

For most ``alternate backends'' (see Section \ref{sec:alternate_backends} and Table \ref{table:backends}), where scaling the intensities or passing passband luminosities directly is not possible, the returned fluxes or magnitudes need to be re-scaled into the units used by PHOEBE by estimating (using the same assumptions as above) the total passband flux from the individual passband luminosities:  

\begin{equation}\label{eq:F_backend_other}
    F_\mathrm{alt.~backend,~scaled} = F_\mathrm{alt.~backend} \sum_\mathrm{stars} \frac{L_{\mathrm{pb, rel}|w}}{4 \pi}
\end{equation}

When using the \textit{``decoupled''} mode, the desired luminosities, $L_{\mathrm{pb, rel}|w}$ per-star can be used directly to estimate this flux scaling without the need for computing absolute luminosities.  But for any other mode, one or more relative passband luminosities must be computed internally (via Equation \ref{eq:S_rel}) by first determining the absolute luminosities of each component.  As PHOEBE Legacy follows similar logic to PHOEBE 2 and takes passband luminosities as input; the respective $L_{\mathrm{pb, rel}|w}$ are passed directly and the returned fluxes do not need to be rescaled with Equation (\ref{eq:F_backend_other}).  However, unless using \textit{``decoupled''}, \textit{``component-coupled''} (which is supported natively by PHOEBE Legacy), or \textit{``dataset-scaled''}, the absolute luminosites still need to be estimated to handle the appropriate coupling of the relative luminosities.  In any of these cases, two methods are introduced in PHOEBE 2.3 to estimate these absolute luminosities: 

\begin{description}
    \item[PHOEBE meshes at $t_0$] a PHOEBE mesh is created at time $t_0$ using the Roche model for surface distortion and is then populated with the appropriate intensities from the given atmosphere and passband tables and integrated over the entire surface to calculate the intrinsic passband luminosity (Equation \ref{eq:L_pb_abs_phoebe}).  As building the mesh is one of the more computationally expensive steps in PHOEBE, estimating luminosities in this way can largely negate the speed benefits of using these ``alternate backends'', but is more accurate for any significant surface distortion.

    \item[Stefan-Boltzmann approximation] the mean stellar values for $R_\mathrm{equiv}$, $T_\mathrm{eff}$, logg, and abundance are used to query the selected atmosphere and passband tables for the ``mean'' $\langle I_{\perp|, \mathrm{abs}|w} \rangle$ and the integral of the limb-darkening model $\mathcal{D}_\mathrm{int}$.  The absolute passband luminosity for each star is then approximated as:
    
    \begin{equation}
    L_{\mathrm{pb, abs}|w} = 4 \pi R_\mathrm{equiv}^2 \langle I_{\perp,\mathrm{abs}|w} (T_\mathrm{eff}, \mathrm{logg}, \mathrm{abun})\rangle \mathcal{D}_\mathrm{int}(T_\mathrm{eff}, \mathrm{logg}, \mathrm{abun}, \mathrm{ld}) \mathcal P_\mathrm{int|w}
    \end{equation}
    This essentially treats each star as a uniform sphere for the purposes of computing the absolute luminosity (and therefore the scaling factor).  For cases where distortion is minimal or the exact scaling is not important, using this approximation will be significantly faster than requiring a mesh to be built.

\end{description}

It is important to note that, with either method, Equation (\ref{eq:F_backend_other}) makes assumptions in estimating passband flux levels from these luminosities, and so the scaled fluxes cannot be exact.  Furthermore, for backends which accept surface brightness ratio instead of absolute effective temperatures, these are estimated as the ratio of the calculated passband luminosities over the square of their respective equivalent radii.  And lastly, the normalizations natively used by these codes differ slightly, both from each other and from PHOEBE.  \texttt{\sc ellc} normalizes the exposed light curves by the integrated flux over the irradiated surfaces of each component \citep{maxted2016}, whereas \texttt{\sc jktebop} normalizes to the magnitude of the system out-of-eclipse at quadrature (John Southworth, private communication).  The effect of these different treatments on PHOEBE's rescaling can be seen in Figure \ref{fig:backends_compare}.  Despite these drawbacks, this implementation allows for simple support of passband-level effects that may not be supported natively by these other codes and allow for an easier drop-in replacement. 

To be consistent, the fluxes returned from all backends do not include contributions from third-light and distance (they are disabled even for codes that would otherwise natively support these effects to ensure consistency).  In the case where third light is provided in fractional (instead of flux) units, we first convert to flux units from the scaled luminosities of each star:

\begin{equation}\label{eq:l3_frac_flux}
    l3_\mathrm{flux} = \frac{l3_\mathrm{frac}}{1-l3_\mathrm{frac}} \sum_\mathrm{stars} \frac{L_{\mathrm{pb, rel}|w}}{4 \pi}
\end{equation}
where ``stars'' are again the components in the system, not including the source of the extraneous third light.  This uses the same uniform and spherical approximation as was used for the luminosity to flux translation in Equation (\ref{eq:F_backend_other}) and also requires absolute luminosities to be estimated through either of the two methods mentioned above whenever not using the \textit{``decoupled''} mode.

If using the \textit{``dataset-scaled''} mode, the fluxes are instead scaled to the observations -- before the inclusion of third light and so acting directly on absolute fluxes (i.e.~$L_{\mathrm{pb, rel}|w} \equiv L_{\mathrm{pb, abs}|w}$) as passband luminosity scaling and distance are arbitrary in this case -- by determining a flux-scale factor, $S_\mathrm{ds}$, via least squares:

\begin{equation}\label{eq:ds_l3_flux}
    F_\mathrm{synthetic,~dataset-scaled} = S_\mathrm{ds} F_\mathrm{backend} + l3_\mathrm{flux}
\end{equation}
For any case where $l3_\mathrm{frac}$ is provided instead, the conversion to $l3_\mathrm{flux}$ will also include this same scale factor:

\begin{equation}\label{eq:ds_l3_frac}
    F_\mathrm{synthetic,~dataset-scaled} = S_\mathrm{ds} \left( F_\mathrm{backend} + \frac{l3_\mathrm{frac}}{1-l3_\mathrm{frac}} \sum_\mathrm{stars} \frac{L_{\mathrm{pb, abs}|w}}{4 \pi} \right)
\end{equation}
PHOEBE 2.3 also introduces support for using \textit{``dataset-scaled''} on multiple light curves which are coupled together using the \textit{``dataset-coupled''} mode.  In this case, Equations (\ref{eq:ds_l3_flux}) and (\ref{eq:ds_l3_frac}) are used, as appropriate, but fitting for a single $S_\mathrm{ds}$ for all the coupled data sets simultaneously.

In all other (non-\textit{``dataset-scaled''}) modes, distance and third light are then included on top of the returned fluxes from the backend:

\begin{equation}
    F_\mathrm{synthetic} = \frac{F_\mathrm{backend}}{d^2} + l3_\mathrm{flux}
\end{equation}
where $F_\mathrm{backend}$ is either adopted from PHOEBE (Equation \ref{eq:F_backend_phoebe}) or any alternate backend after rescaling, if necessary (Equation \ref{eq:F_backend_other}).

\subsubsection{Systemic Velocity and Per-Component Radial Velocity Offsets}\label{sec:rv_offset}

Similarly to distance and third light being included on top of the respective backend fluxes, radial velocities directly from any backend exclude any offsets (even for backends that do natively support systemic velocities).  The final synthetic radial velocities are then defined as:

\begin{equation}
    RV_\mathrm{synthetic} = RV_\mathrm{backend} + RV_\gamma + RV_\mathrm{offset}
\end{equation}
where $RV_\gamma$ is the constant barycentric systemic velocity added to all RVs (for all components across all data sets), whereas $RV_\mathrm{offset}$ allows for per-dataset and per-component offsets to the velocities.

These per-component offsets are particularly useful for hot stars, where offsets in spectral lines are commonplace due to being created in different levels of the atmospheres \citep{underhill1994, harmanec2002, shenar2018}.  The magnitude of this offset can vary between the two components in the system but also between radial velocity data sets which were reduced from different spectral lines.  By building this into the model itself, these offsets can be marginalized over and any correlations (with the mass-ratio or systemic velocity $RV_\gamma$, for example) can be properly accounted for in the posteriors.

\subsection{Noise Nuisance}\label{sec:noise_nuisance}

Determining posteriors and uncertainties of physical parameters of the system depends strongly on the observational uncertainties.  Unfortunately, these are often known to be underestimated, so it is important that any such underestimation does not propagate through to a biased solution or underestimated model uncertainties.

In the cases where the quoted observational uncertainties may be underestimated by a constant factor (such as in \textsl{Kepler} data; cf.~\citealt{jenkins2017}), we can introduce the uncertainty scaling factor directly into observational uncertainties in the definition of the merit function (see also Equations \ref{eq:chi2} and \ref{eq:chi2sigma}):
    \begin{equation}
        \sigma^2 = \sigma_\mathrm{o}^2 + y_{m}^2  \mathrm{e}^{(2  \sigma_\mathrm{lnf})}.
    \end{equation}

This can be particularly useful as a ``nuisance parameter'' while sampling to determine posteriors (see Sections \ref{sec:samplers} and \ref{sec:posteriors}).  By marginalizing over this underestimation factor, any degeneracies between the factor itself and the physical model parameters can be encoded in the resulting posteriors\footnote{See the example for fitting a line to data with underestimated uncertainties in the \texttt{\sc emcee} online documentation at \url{https://emcee.readthedocs.io/en/v3.0.0/tutorials/line/}.}.

We refer the reader to \citet{hogg2010}, which provides a practical overview to handling observational uncertainties in several different situations.

\subsection{Gaussian Processes}\label{sec:gaussian_processes}

A Gaussian process (GP) is a random process in which each data point is drawn from the assigned random variable where the joint distribution of all variables is Gaussian.  GPs allow modeling the noise correlations (such as serial correlation and/or heteroscedasticity) on top of the astrophysical signal described by the noise-free forward-model, and thus replace the need to fit functions (such as polynomials) to the data for the purpose of detrending. Within PHOEBE, GPs are applied to the forward-model and, as such, are available for any of the supported computation backends with any choice of the kernel (which describes the covariances between adjacent points). PHOEBE implements GPs via \texttt{\sc celerite}: a Python implementation of GPs that aims to be computationally efficient and scalable to a large number of data points \citep{celerite}.  PHOEBE 2.3 supports any combination of Matern $3/2$ and Simple Harmonic Oscillator kernels, but intentionally excludes the white noise term (``jitter'') natively supported by \texttt{\sc celerite} to avoid conflicting with the noise nuisance parameter separately implemented within PHOEBE (Section \ref{sec:noise_nuisance}).

After the original forward-model is computed and interpolated onto the observation times (if necessary), the residuals between the forward-model and observations are determined.  These residuals and the input parameters for the kernels are then passed to \texttt{\sc celerite}, which then provides the GP component of the model.  The forward model and GP contribution are then added, resulting in the final model (which in turn is used to compute the residuals or merit function for optimizers or samplers).  Note that GPs require the final forward-model to be exposed at the exact times of the observations although the physical forward model is still computed at the requested times or phases as discussed in Section \ref{sec:solver_times}.

\begin{figure}
    \subfloat{%
      \includegraphics[width=0.5\columnwidth]{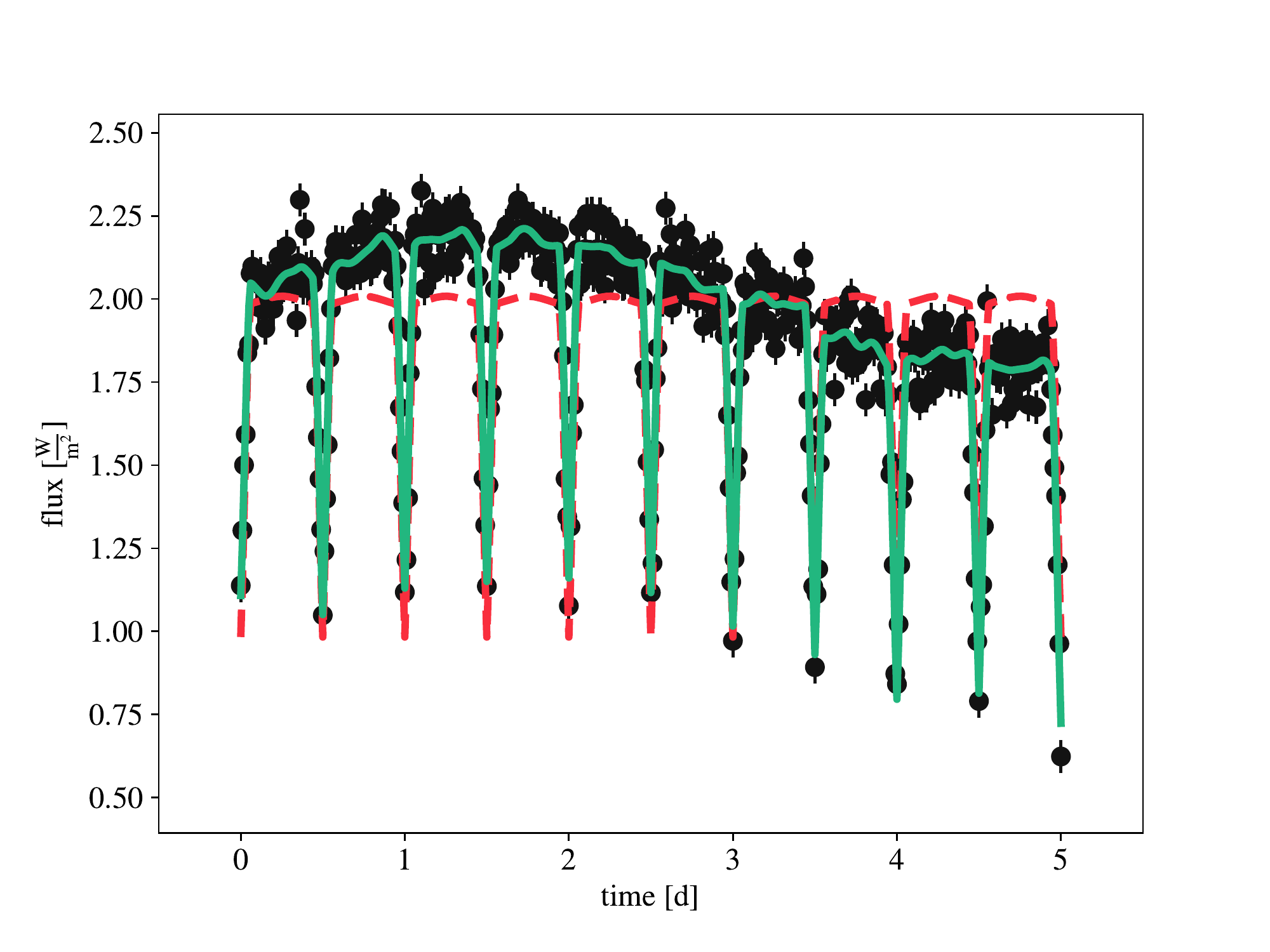}
    }
    \subfloat{%
      \includegraphics[width=0.5\columnwidth]{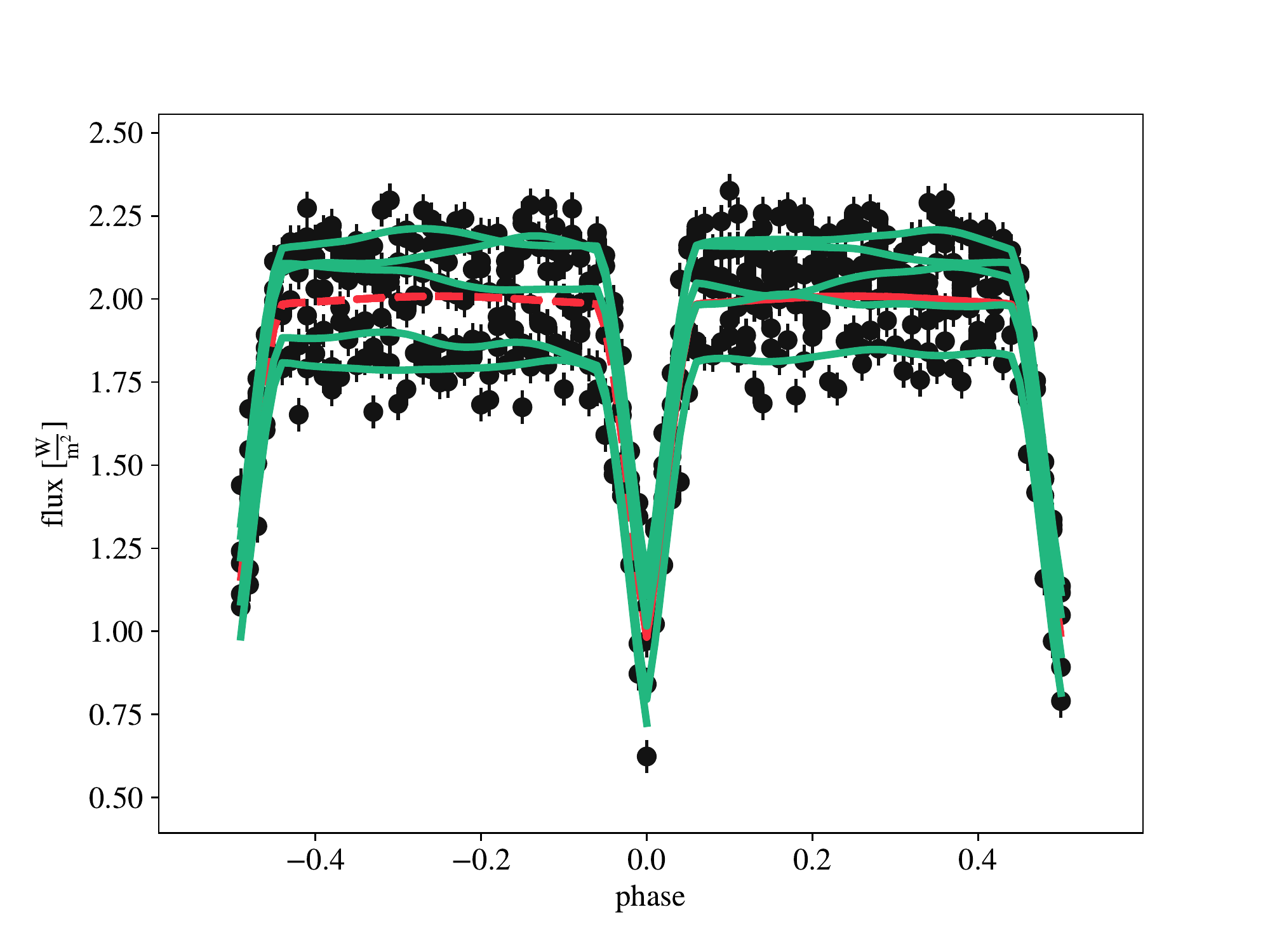}
    }
    \caption{Forward-model both without (red dashed lines) and with (green solid lines) the Gaussian processes. Two Gaussian process kernels (a Matern 3/2 kernel and a simple harmonic oscillation kernel) are shown in time-space (left) and phase-space (right).}
    \label{fig:GPs}
\end{figure}

Figure \ref{fig:GPs} showcases an example of including synthetic observations with additional superimposed trends and noise.  Here the inclusion of GPs accounts for the residuals between the observations and the synthetic forward model.

While GPs would ideally account only for the noise component (stochastic \emph{and} correlated), their high degree of freedom can lead to overfitting the data and accounting for some of the signal. This can manifest in excess GP spectral power at the orbital period and, in consequence, it leads to degenerate solutions and it affects parameter posteriors. It is thus important to properly marginalize (see Section \ref{sec:samplers}) over GP parameters to avoid, or at least quantify, the level of overfitting.

\subsection{Parallelization}

The PHOEBE backend itself is parallelized at the per-time level via a Message Passing Interface (MPI) implementation. Here the initial setup is done by a single processor which then sends individual time stamps to the remaining processors to compute the requested synthetic observables at those times.  The main processor then compiles and orders the results and exposes the model. The wrappers around other backends generally support parallelization at the per-dataset level -- allowing each processor to compute the observables for all times, but for a single data set.  However, within optimizers or samplers that support that capability, the parallelization within the forward-models is disabled in favor of parallelizing at the per-model level via either MPI or multiprocessing.  As this has less overhead, it is often more efficient.

It should be noted that for Markov Chain Monte Carlo (MCMC; see Section \ref{sec:mcmc}), for example, this loses its advantages if there are more processors available than requested walkers.  Whenever this is the case, PHOEBE automatically switches to run MCMC in serial with each iteration parallelized per-time.

In addition, if not using MPI, PHOEBE will make use of a multiprocessing pool across all available processors for any backend that supports multiprocessing (currently emcee and dynesty).

\section{Optimizers}\label{sec:optimizers}

PHOEBE 2.3 includes wrappers around several optimization algorithms from \texttt{scipy.optimize}\footnote{PHOEBE 2.3 requires \texttt{scipy} 1.2+.} \citep{2020SciPy-NMeth}, including: Nelder-Mead \citep{nelder-mead}, Powell, and conjugate gradient.  These optimizers can be quite efficient at improving a model fit once already in the vicinity of the optimal solution -- found via estimators or a full global parameter space search, for example.   

PHOEBE allows for choosing which parameters to be adjusted by the optimizer as well as the ability to define priors, if desired.  The respective algorithm is then called by passing the negative log-likelihood (see Section \ref{sec:merit_function}) to the appropriate \texttt{\sc scipy} algorithm.  The optimization results in the maximum a priori (MAP) solution if priors are defined, or the maximum likelihood estimation (MLE) otherwise.  It is important to note that, irrespective of whether the user provides priors or not, the parameter limits, wrapping limits, and any failed models still act as uninformative priors which penalize the merit function.

\begin{figure}
    \centering
    \includegraphics[width=0.5\textwidth]{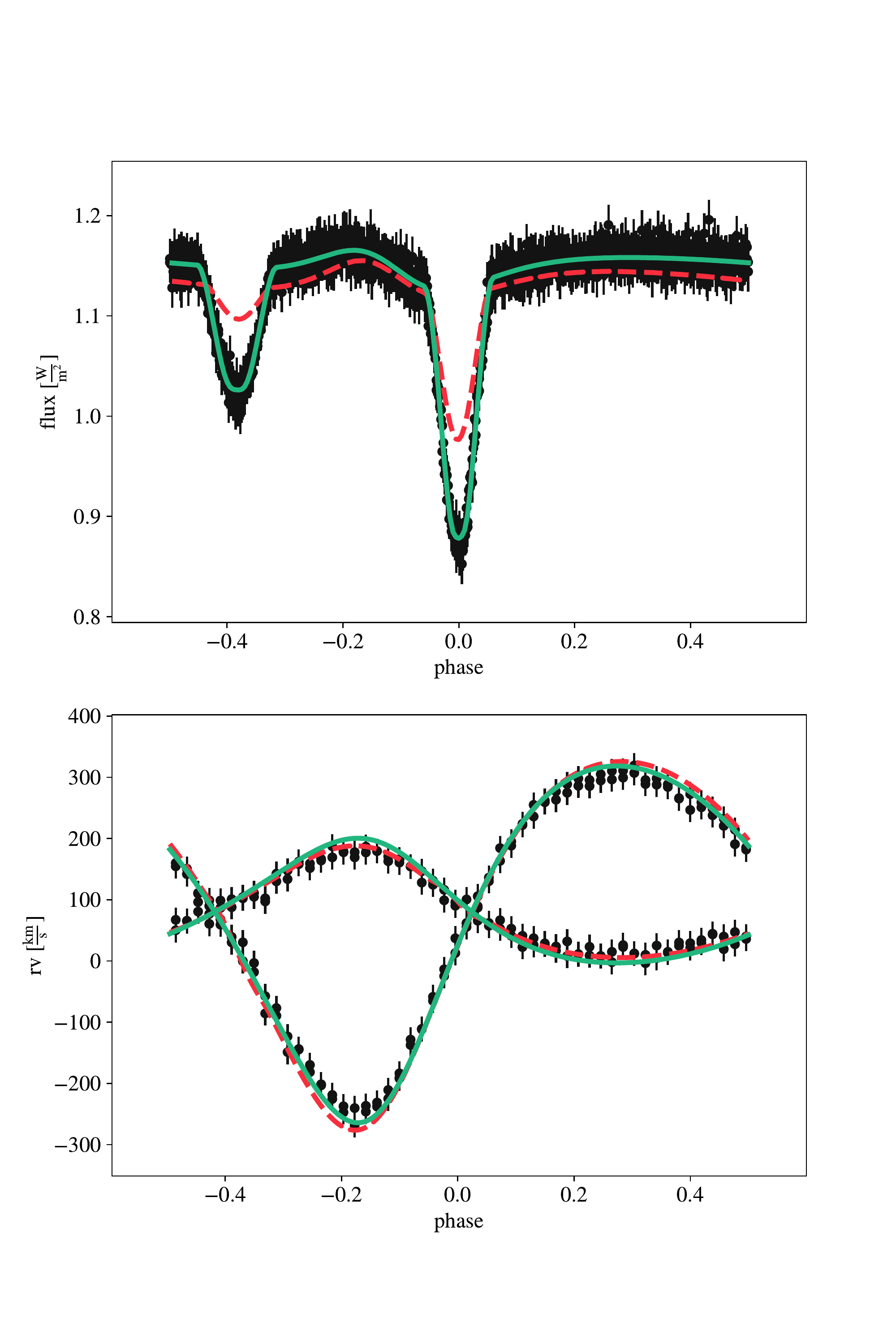}
    \caption{Model improvement after adopting proposals from estimators (dashed red lines) and after running Nelder-Mead (solid green lines).}
    \label{fig:optimizer_nm}
\end{figure}

As with estimators (Section \ref{sec:estimators}), PHOEBE then exposes the proposed values for each of the adjusted parameters along with the value of the merit function before and after optimization as well as any diagnostic values returned by \texttt{\sc scipy}, allowing the user to decide which, if any, of the proposed values to adopt.  Figure \ref{fig:optimizer_nm} shows the results of a Nelder-Mead optimization on several parameters starting at the proposed values from estimators.  After just a few hundred iterations, the residuals and $\chi^2$ show that the model has improved significantly. 

\section{Samplers}\label{sec:samplers}

Samplers are not designed for finding the global solution, rather, their purpose is to explore the shape of the local parameter space and provide robust posteriors and uncertainties that expose the underlying degeneracies between various parameters.

\subsection{Markov Chain Monte Carlo (emcee)}\label{sec:mcmc}

Markov Chain Monte Carlo (MCMC) is a Bayesian method to sample the parameter space in order to estimate the posterior density function -- including any correlations between the sampled parameters -- by using the log probability as the merit function (see Section \ref{sec:merit_function}).   Affine-Invariant MCMC algorithms  \citep{goodmanweare2010} sample this same space based on the \textit{ensemble} of workers and generally has a shorter burn-in period and requires less user-tuning \citep{foreman-mackey2013}, and is therefore a good candidate to use whenever the forward-model is computationally expensive and the probability space may not be orthogonal.

The \texttt{\sc emcee}\footnote{PHOEBE 2.3 has been tested with \texttt{\sc emcee} v3.0 (and does not support earlier versions).} \citep{foreman-mackey2013, emceev3} python package is an Affine-Invariant Markov Chain Monte Carlo Python implementation that has been widely adopted by the field.  PHOEBE includes a wrapper around \texttt{\sc emcee}, exposing most of the basic functionality along with the convenience to easily add or remove sampled parameters and use complex distributions for priors and the initial sample, while handling the merit function and parallelization issues.  However, as with any algorithm, it is still important to understand \texttt{\sc emcee} itself, how best to adjust the inputs, and how to interpret the results.  \citet{foreman-mackey2013} provides a good overview and discussion and the online documentation for \texttt{\sc emcee} (\url{https://emcee.readthedocs.io}) provides several helpful examples.

\begin{figure}
    \centering
    \includegraphics[width=1.0\textwidth]{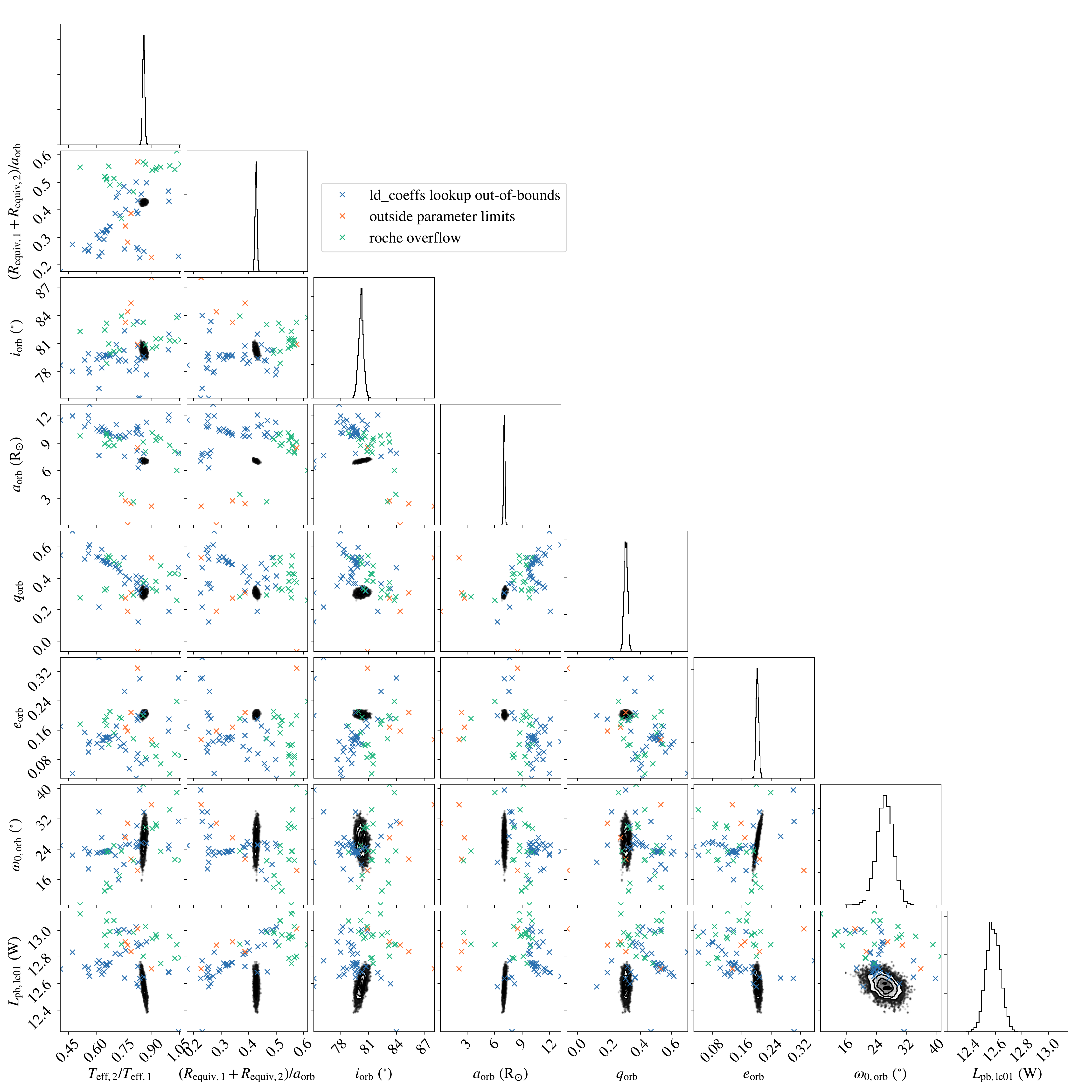}
    \caption{An example corner plot showing the posteriors from an emcee run along with the positions in the parameter space of ``failed/rejected samples''.  The legend shows that some of the values proposed by emcee were rejected due to Roche overflow or a failure to lookup appropriate limb darkening coefficients, for example.  This plot is particularly useful to diagnose any issues, especially if the posteriors seem to be pushing against a boundary.}
    \label{fig:emcee_failed_samples}
\end{figure}

When preparing an \texttt{\sc emcee} run through PHOEBE, the user can define the distributions to be used for the initial sample (which defines which parameters will be included in the samples) in any preferred parameterization and, optionally, distributions for the priors.  If priors are not provided, uninformative distributions are still adopted based on the limits of the individual parameters and angle wrapping limits (see Section \ref{sec:angle_wrapping}).  Furthermore, if any individual model fails for any reason, the merit function (see Section \ref{sec:merit_function}) will return $- \infty$, essentially acting as a prior on the allowed parameter space due to unphysical systems or limitations of the backend itself.  The causes for each of these rejected samples can optionally be exposed to the user, as shown in Figure \ref{fig:emcee_failed_samples}.

Although the initial sample and the priors can be the same set of distributions, it is generally good practice to start \texttt{\sc emcee} in an N-dimensional hyperball (i.e.~via Gaussian distributions on multiple parameters) around the best-known solution from optimization (see Section \ref{sec:optimizers}).  Alternatively, it is possible to sample directly from the priors themselves (although this is generally less efficient and more susceptible to getting stuck in local solutions -- see \citet{foreman-mackey2013} for a discussion on this topic).  Priors, on the other hand, can either be informative -- uncertainties from an external analysis or from the literature -- or uninformative -- conservative uniform distributions to prevent the chains from ``wandering'' into parameter space that is clearly incorrect or unphysical, and are included in the merit function as described in Section \ref{sec:merit_function}. 

In addition to distributions, the user can set options for the number of processors, number of walkers, and number of iterations.  After the run is complete (or at intermediate steps, as requested by the user), the user can view the progress of the chains and the log-probability versus iteration and adjust the burnin and thinning parameters as necessary.  If the run is completed but has not yet converged, PHOEBE allows continuing an \texttt{\sc emcee} run from the existing chains.

In addition to the full chains, the acceptance fractions and autocorrelation times are exposed to the user.  As a rule of thumb, \citet{foreman-mackey2013} suggests acceptance fractions to all be between 0.2 and 0.5 and enough iterations to cover roughly ten autocorrelation times.  PHOEBE then automatically determines defaults for thinning and burn-in based on the autocorrelation times of the chains:  

\begin{equation}
    \mathrm{burnin} = F_\mathrm{burnin} \max( \tau_\mathrm{autocorr})
\end{equation}
and
\begin{equation}
 \mathrm{thin} = F_\mathrm{thin} \min( \tau_\mathrm{autocorr})   
\end{equation}
where $F_\mathrm{burnin}$ and $F_\mathrm{thin}$ are user-defined scaling factors that default to 2 and 0.5, respectively, and $\tau_\mathrm{autocorr}$ are the estimated autocorrelation times, per-parameter, as exposed by \texttt{\sc emcee}.

Once the chains sufficiently cover the parameter space and are considered to be converged, the user can then adjust the thinning, burn-in, and/or a cutoff in log-probability to apply to the chains returned by \texttt{\sc emcee} and generate a multivariate distribution of the resulting posteriors (see more in Section \ref{sec:posteriors}).

\subsection{Dynamic Nested Sampling (dynesty)}\label{sec:nested_sampling}

Nested sampling is another Bayesian method for sampling the parameter space.  Unlike MCMC, which samples from an initial distribution, walks the parameter space, and is ``penalized'' by the priors in the merit function, nested sampling continually samples directly from the priors themselves which are broken into slices \citep[see ][for an in-depth comparison and discussion]{speagle2020}.  Nested sampling, unlike MCMC, is capable of exploring and exposing multi-modal posteriors.  However, with wide uninformative priors, nested sampling can become prohibitively expensive.  \textit{Dynamic} nested sampling somewhat alleviates this expense by dynamically changing the number of samples per iteration as the algorithm converges to the final posterior.  Additionally, unlike MCMC which is allowed to wander outside the initial sampling distributions as long as the priors allow it, dynamic nested sampling will never consider any solution outside the original parameter-space defined by the priors.

The \texttt{\sc dynesty}\footnote{PHOEBE 2.3 has been tested with \texttt{\sc dynesty} v1.0.} code is a dynamic nested sampling Python package.  As with \texttt{\sc emcee}, it is very useful to first understand the details and intricacies of dynamic nested sampling and the \texttt{\sc dynesty} code.  \citet{speagle2020} provides detailed discussion and the online documentation (\url{https://dynesty.readthedocs.io}) gives numerous examples.

Since \texttt{\sc dynesty} samples directly from the priors, PHOEBE excludes the priors from the log-likelihood.  Instead, PHOEBE makes use of \texttt{\sc distl} (see Section \ref{sec:priors}) to transform the defined priors into prior transforms which map a random value between 0 and 1 onto the drawn values of each individual parameter, and passes these to \texttt{\sc dynesty}.  It is important to note that this transformation requires ignoring any possible covariances between parameters in the priors (i.e.~if priors were provided as multivariate distributions they would be first flattened into univariate distributions before being passed to \texttt{\sc dynesty}).

After calling \texttt{\sc dynesty} from within PHOEBE, the user has access to all of the output arrays from \texttt{\sc dynesty} itself to access or plot any diagnostic figures.  The samples from \texttt{\sc dynesty} can then be transformed into posterior distributions by accounting for their respective weights.

\section{Posteriors \& Uncertainties}\label{sec:posteriors}

\begin{figure}
    \subfloat{%
      \includegraphics[width=0.5\columnwidth]{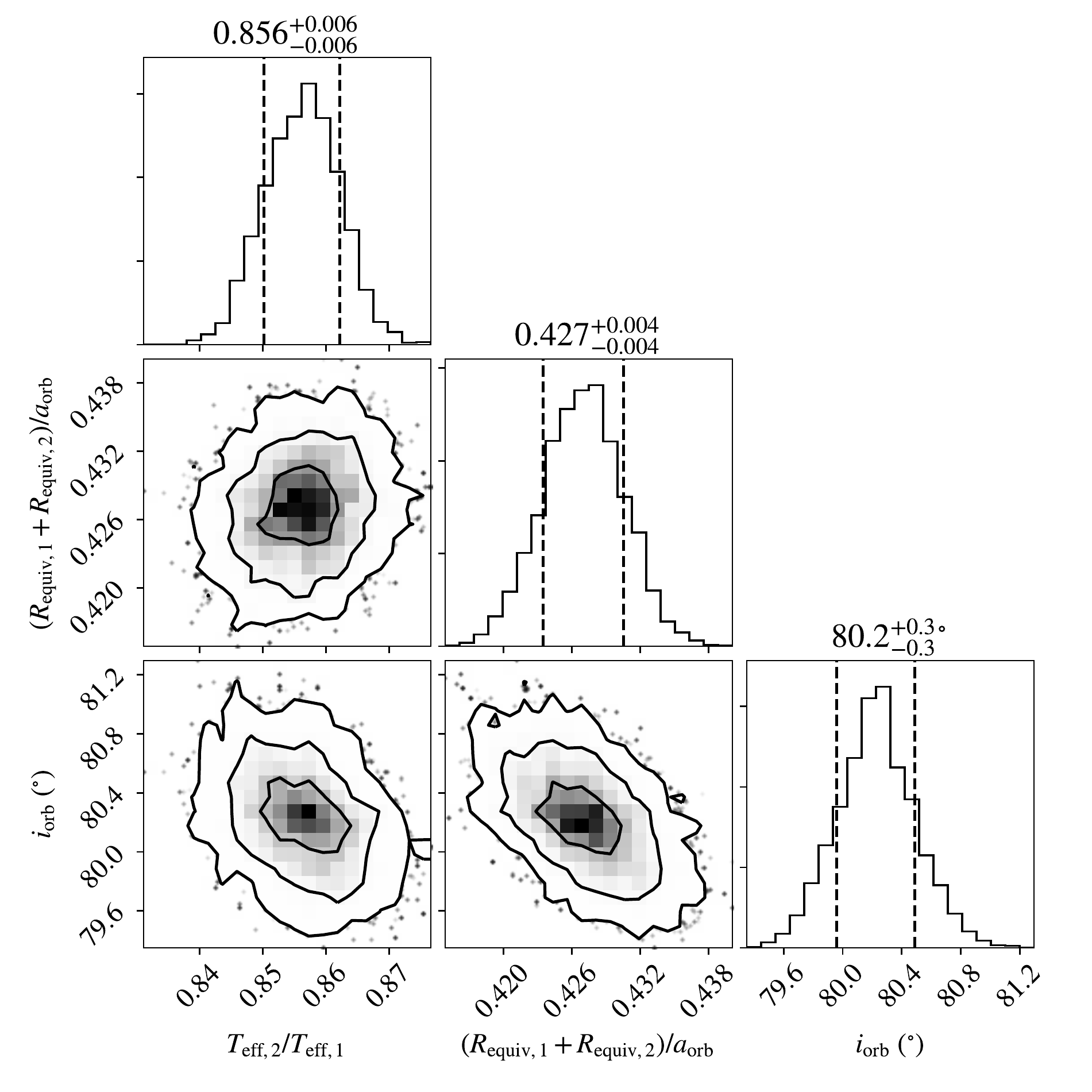}
    }
    \subfloat{%
      \includegraphics[width=0.5\columnwidth]{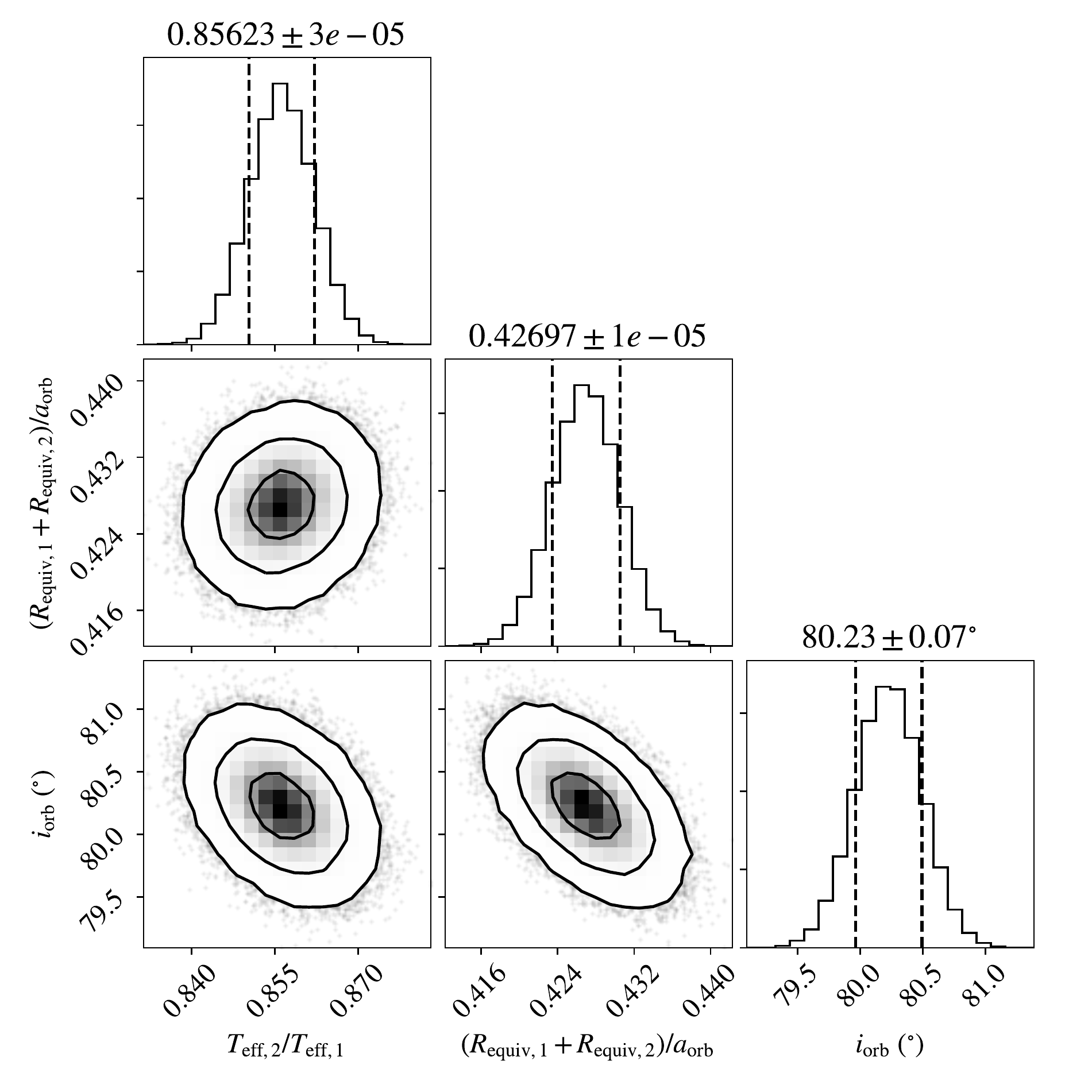}
    }
    \caption{An example corner plot of the (burn-in and thinning applied) posteriors directly from emcee results (left).  These can optionally be translated into a multivariate guassian distribution (right) which, when appropriate, represents the same information with only the means and covariance matrix.  This is particularly useful to include the means and covariances in a publication or to reuse the posteriors as priors.}
    \label{fig:posteriors_mvsamples_mvgaussian}
\end{figure}

The results from any sampler can be accessed as a posterior distribution -- representing not only the uncertainties on each of the sampled parameter values, but also the correlations (or covariances) between all of the sampled parameters.  When these distributions are sufficiently gaussian, the posteriors can easily be converted into a multivariate gaussian distribution which represents the posteriors as just the means and a covariance matrix.  Figure \ref{fig:posteriors_mvsamples_mvgaussian} shows an example corner plot of the posteriors of several parameters obtained from \texttt{\sc emcee} samples and their conversion into a multivariate gaussian.

\begin{figure}
    \subfloat{%
      \includegraphics[width=0.5\columnwidth]{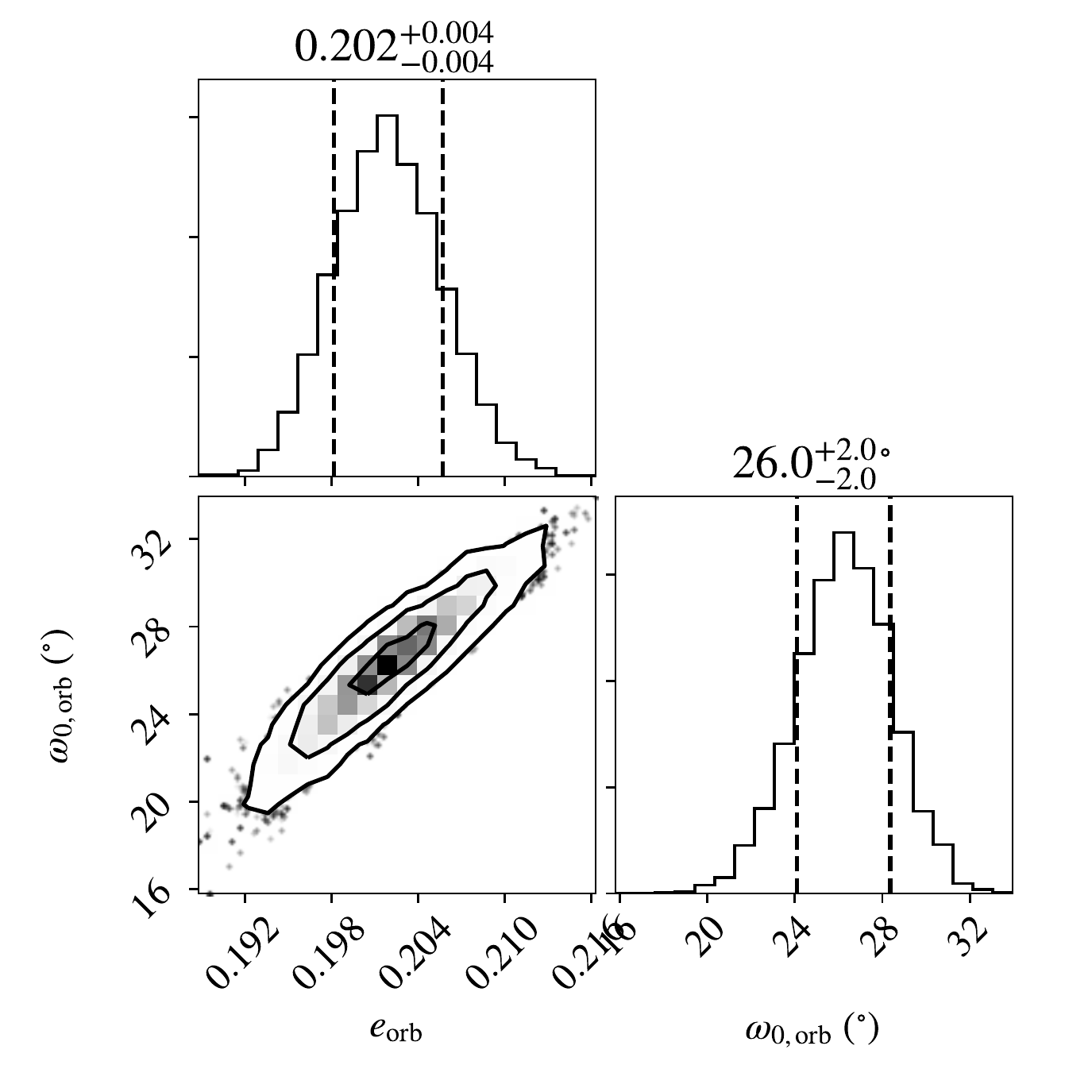}
    }
    \subfloat{%
      \includegraphics[width=0.5\columnwidth]{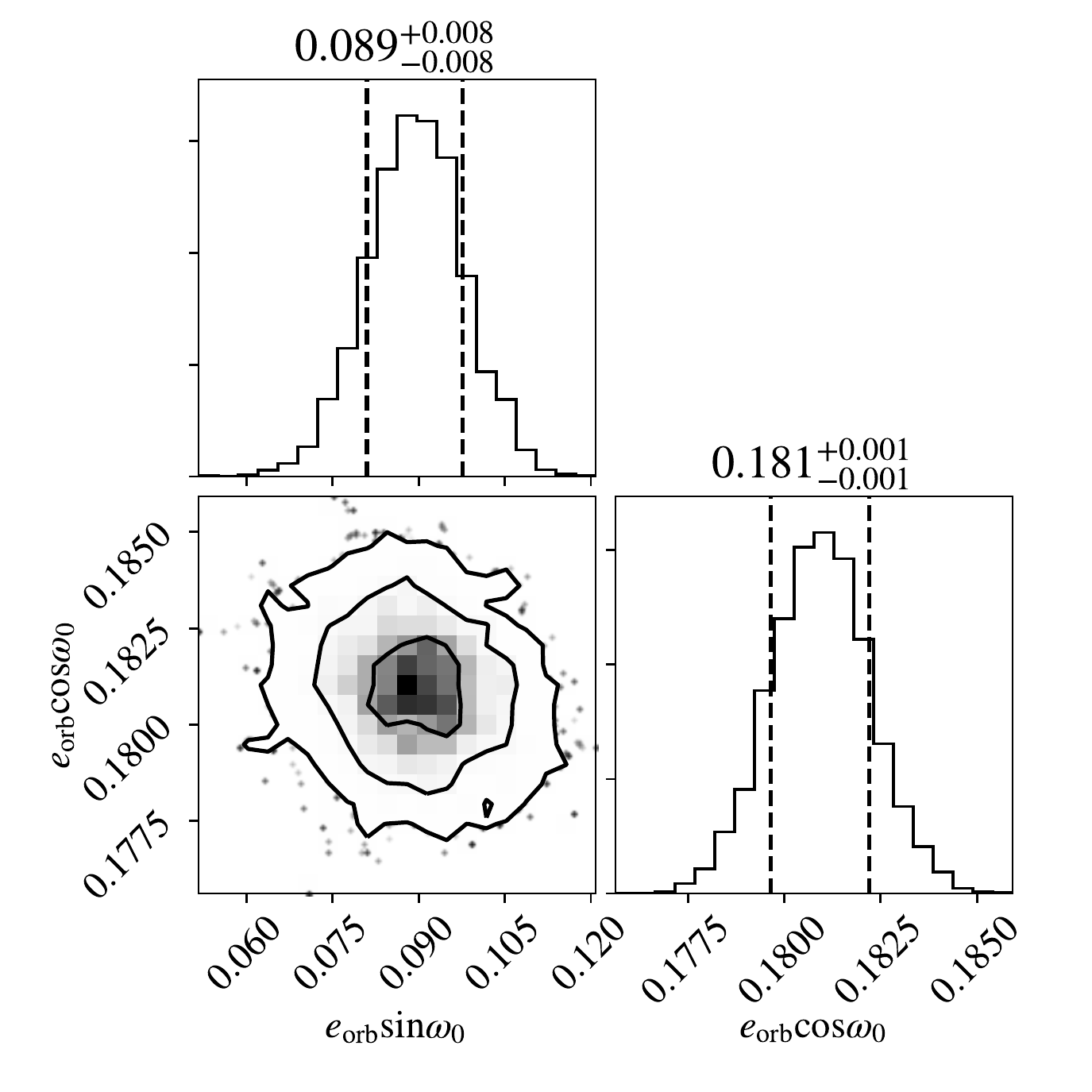}
    }
    \caption{An example corner plot of the eccentricity and argument of periastron posteriors from an emcee run (left).  Although the distributions are correct, these are by nature both non-gaussian and highly correlated, making it difficult to parameterize the results for inclusion in a publication, for example.  These posteriors can be propagated through the respective constraints (see Section \ref{sec:parameterization}) to instead expose the posteriors in $e \sin \omega$ and $e \cos \omega$ which are more orthogonal and can often conveniently be converted to gaussian posteriors.}
    \label{fig:posteriors_parameterization}
\end{figure}

As these distributions are \texttt{\sc distl} distribution objects, they can easily be manipulated into any of the supported distribution types, saved to a file, plotted, expose symmetric or asymmetric uncertainties to include in published results, or used for further sampling.  For example, posteriors from one sampling run can be adopted and used as either the initial sampling distribution or priors for another sampling run.  This can be useful to optimize and determine posteriors for one data set, but maintain that information in the log-likelihood for another data set or when new observations become available.  As with distributions used for priors (Section \ref{sec:priors}), these posteriors can be propagated through the ``constraints'' and exposed in any desired parameterization (Section \ref{sec:parameterization}).  For example, Figure \ref{fig:posteriors_parameterization} shows the \texttt{\sc emcee} samples for eccentricity and argument of periastron which can then be propagated to $e \sin \omega_0$ and $e \cos \omega_0$ which are easier to represent as a multivariate gaussian and report in published scientific results.  This allows for independent choices for the combination of parameters for sampling, priors, and posteriors, giving the full flexibility needed based on the available observations and known information.

It is important to note that the degeneracies and uncertainties determined by any sampler are computed under the assumption that all remaining parameters are held fixed at their face values with no uncertainties, leading to uncertainty underestimation or possibly even an incorrect solution.  For example, if the value of the orbital period is believed to be well-known and therefore left fixed during sampling, the resulting posteriors -- and therefore uncertainties -- on all other sampled parameters will exclude any degeneracies with respect to the orbital period, and will therefore likely be underestimated.  Likewise, in the case where the assumed fixed value is incorrect, the remaining sampled parameters may be incorrect as well.  In effect, all parameters that are not sampled are treated as if they have a delta function for a prior and are known to infinite precision. In a perfect world, all parameters would be sampled or marginalized over, but practically that is not feasible with tens or even hundreds of possible parameters.  

\begin{figure}
    \subfloat{%
      \includegraphics[width=0.5\columnwidth]{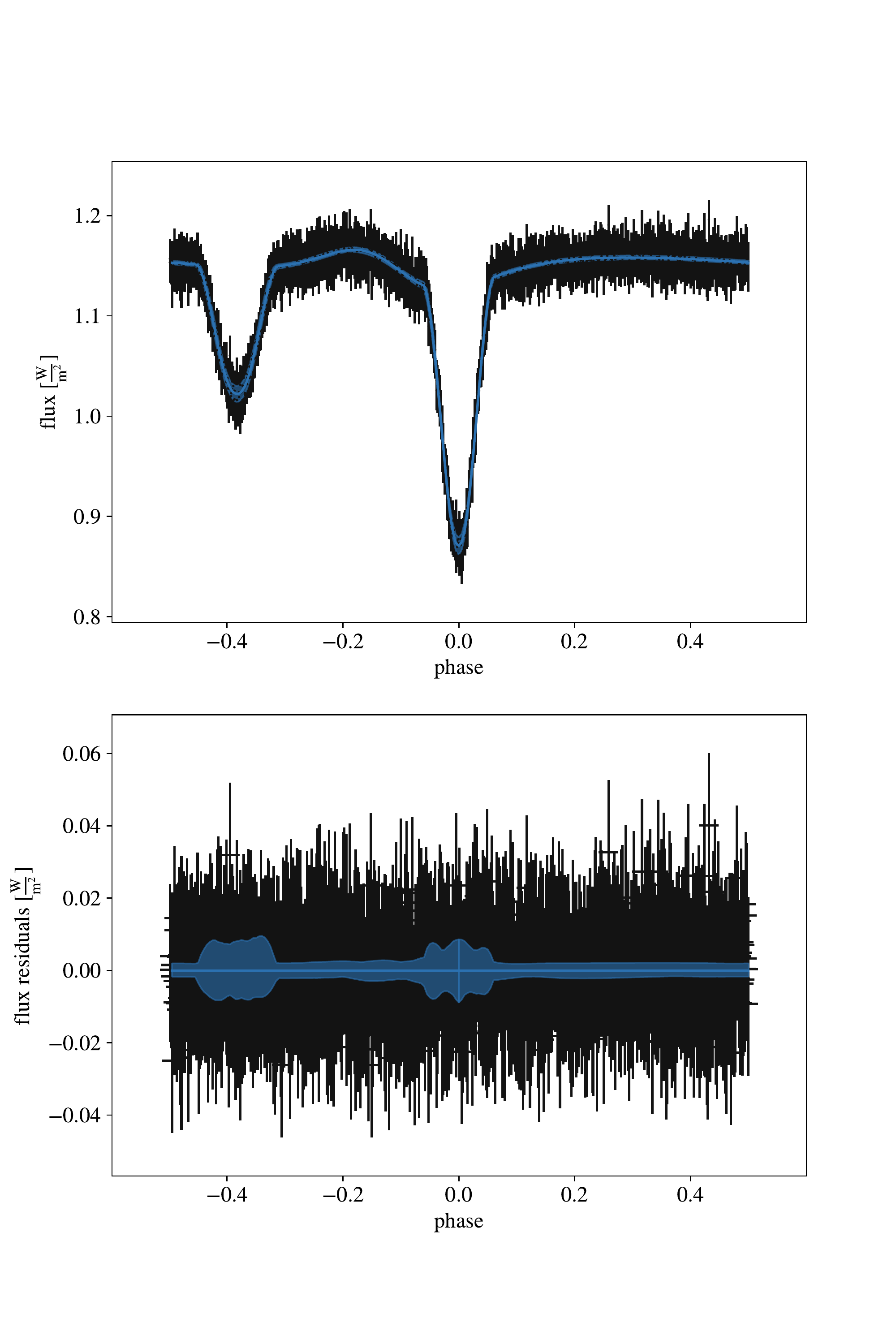}
    }
    \subfloat{%
      \includegraphics[width=0.5\columnwidth]{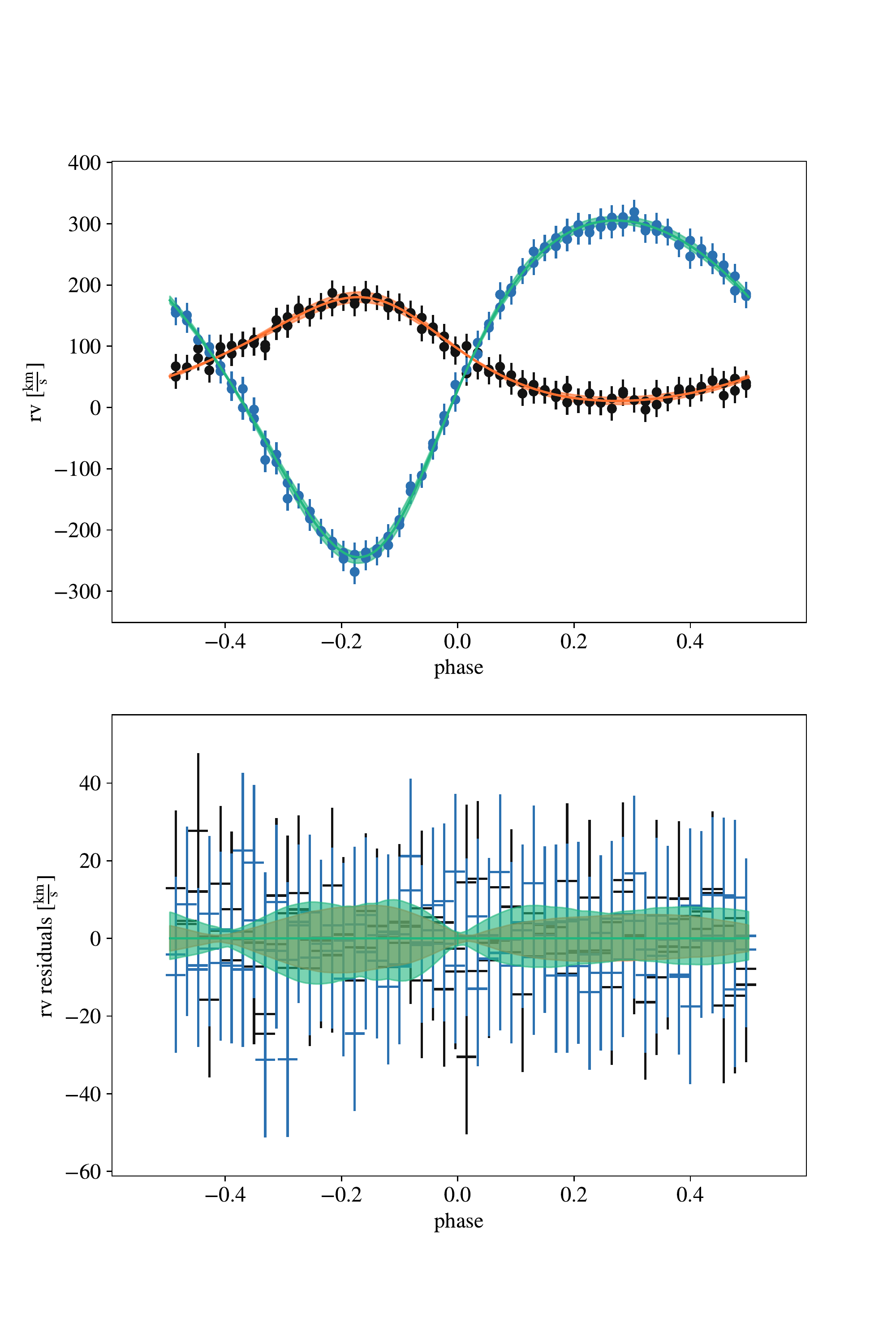}
    }
    \caption{Any distribution (in this case the posteriors from the emcee run) can be propagated through the forward-model.  Here the three-sigma uncertainties on the light curve (left) and radial velocity curves (right) and their residuals (bottom) are shown, depicting that the spread in the observations, given their uncertainties, are well represented by the spread in the model caused by the uncertainties in the posteriors.}
    \label{fig:posteriors_sample_from}
\end{figure}

Lastly, as with any distribution, PHOEBE now allows propagating posteriors through the forward-model itself.  By running the forward-model over a number of samples from the posterior distributions, PHOEBE can compute and expose these individual models or the median model and 1-, 3-, or 5- sigma spreads in flux or radial velocity.  Figure \ref{fig:posteriors_sample_from} shows an example of the 3-sigma spread in the synthetic model when drawn from the posteriors from an \texttt{\sc emcee} run.

\section{User Interface}\label{sec:ui}

The PHOEBE 2.3 release also introduces a user interface, capable of all of the functionality of both the forward-model and inverse problem described above, including basic plotting functionality, and is available for download at \url{http://phoebe-project.org/clients}.  The user interface is designed with a server--client model, splitting the codebase into three distinct roles: the Python package itself, the user interface client, and a lightweight webserver which can either run on the same machine or remotely from the client and is responsible for running commands sent from the user interface client via the Python package.  This design allows for the flexibility of installing the server on a remote machine with more resources and either installing the user interface locally or accessing a hosted version through the web browser\footnote{a publicly available version is currently hosted at \url{http://ui.phoebe-project.org}.}.

With or without the user interface, PHOEBE also allows the user to conveniently export the forward-model or inverse-problem jobs into a standalone script, which can then be run on a high performance cluster, and the resulting file can be imported locally to view the progress or final results.  This allows for running all interactive and visual components of the fitting process on a local machine while offloading computationally expensive tasks to a remote machine that may not be capable of hosting a webserver or plotting graphics.

\section{Conclusion}

The 2.3 release of PHOEBE builds on past releases and introduces a general framework for data fitting running several common algorithms to address the inverse problem, and an interface for defining and dealing with complex distributions. It also introduces the ability to run the forward-model through several other publicly available codes in addition to PHOEBE, and a web-based user interface.

Although far from a black-box automated fitting pipeline, this common interface aims to ease the learning curve and effort required to employ multiple algorithms and codes while attempting to obtain both accurate and precise parameter values and uncertainties for eclipsing binary systems.  As additional algorithms and forward-models continue to be adopted by the field, we plan to incorporate them into PHOEBE within this same framework.

PHOEBE is an open-source project under the GPL3 license and is hosted at \url{http://phoebe-project.org} and \url{https://github.com/phoebe-project/phoebe2}.  Contributions and feedback are welcome.

\acknowledgments

The development of PHOEBE is possible through the NSF AAG grants \#1517474 and \#1909109 and NASA 17-ADAP17-68, which we gratefully acknowledge.

We thank John Southworth and Pierre Maxted for their permission and discussions regarding their codes, \texttt{\sc jktebop} and \texttt{\sc ellc}.

DJ acknowledges support from the State Research Agency (AEI) of the Spanish Ministry of Science, Innovation and Universities (MCIU) and the European Regional Development Fund (FEDER) under grant AYA2017-83383-P.  DJ also acknowledges support under grant P/308614 financed by funds transferred from the Spanish Ministry of Science, Innovation and Universities, charged to the General State Budgets and with funds transferred from the General Budgets of the Autonomous Community of the Canary Islands by the Ministry of Economy, Industry, Trade and Knowledge.

KH gratefully acknowledges support from NASA ADAP grant 18-ADAP18-228.

DRH gratefully acknowledges the support of the Australian Government Research Training Program.

MA acknowledges support from the FWO-Odysseus program under project G0F8H6N.

\vspace{5mm}

\software{  PHOEBE \citep{prsa2016, horvat2018, jones2020},
            distl (\url{https://github.com/kecnry/distl}),
            EBAI \citep{prsa2008, 2019ascl.soft08018P},
            ellc \citep{maxted2016},
            jktebop \citep{southworth2004, southworth2007, southworth2009, southworth2011},
            celerite \citep{celerite},
            emcee \citep{foreman-mackey2013, emceev3},
            dynesty \citep{speagle2020},
            schwimmbad \citep{2017JOSS....2..357P},
            numpy   \citep{numpy},
            scipy \citep{2020SciPy-NMeth},
            astropy \citep{2013A&A...558A..33A},
            matplotlib \citep{2007CSE.....9...90H},
            corner \citep{corner}
          }
          
\appendix
\section{Description of Symbols}\label{app:symbols}

\begin{description}
    \item [$a_\mathrm{orb}$]: orbital semi-major axis.
    \item [$a_\mathrm{orb} \sin i$]: orbital semi-major axis projected along the line-of-sight.
    \item [$a_\mathrm{comp} \sin i$]: component semi-major axis projected along the line-of-sight.
    \item [$A_v$, $R_v$, $E(B-V)$]: extinction values.
    \item [$e$]: orbital eccentricity.
    \item [$f_\mathrm{orb}$, $f_\mathrm{rot}$]: orbital and rotational frequencies, respectively.
    \item [$i$]: orbital inclination.
    \item [$P_\mathrm{orb}$, $\dot{P}_\mathrm{orb}$]: orbital period at time $t_0$ and its time derivative, respectively.
    \item [$q$]: mass ratio defined as $M_2/M_1$.
    \item [$R_\mathrm{equiv}$]: stellar equivalent (volumetric) radius.
    \item [$t_{0, \mathrm{supconj}}$]: reference time of superior conjunction.
    \item [$T_\mathrm{eff}$]: stellar effective temperature.
    \item [$v_\gamma$]: systemic velocity.
    \item [$\omega_0$, $\dot{\omega}$]: argument of periastron at time $t_0$ and its time derivative, respectively.
    \item [$\Omega_\mathrm{orb}$]: longitude of the ascending node.

\end{description}

\bibliography{refs}


\end{document}